\documentclass[preprint2]{aastex6}

\newcommand{\days}{\mathrm{days}}

\newcommand{\msini}{\ensuremath{m \sin i}}
\newcommand{\mplsini}{\ensuremath{\mpl\sin i}}
\newcommand{\teff}{\ensuremath{T_{\mathrm{eff}}}}

\newcommand{\logg}{\ensuremath{\log{g}}}

\newcommand{\mh}{\ensuremath{\mathrm{[m/H]}}}

\def\mymathhyphen{{\hbox{-}}}
\newcommand{\pvalue}{\ensuremath{p\mymathhyphen\mathrm{value}}}   
\newcommand{\au}{\ensuremath{\mathrm{AU}}}
\newcommand{\dbic}{\ensuremath{\Delta\mathrm{BIC}}}

\newcommand{\kelvin}{\mathrm{K}}

\newcommand{\rsun}{\ensuremath{R_\sun}}

\newcommand{\rstar}{\ensuremath{R_\star}}

\newcommand{\mearth}{\ensuremath{M_\earth}}
\newcommand{\searth}{\ensuremath{S_\earth}}

\newcommand{\mpl}{\ensuremath{M_{p}}}

\newcommand{\seff}{\ensuremath{S_{\mathrm{eff}}}}

\newcommand{\mjup}{\ensuremath{M_{\mathrm{J}}}}

\shorttitle{Jupiter Analogs}
\shortauthors{Buchhave et al.}

\begin{document}

\title{Jupiter Analogues Orbit Stars with an Average Metallicity Close to that of the Sun}

\author{
Lars~A.~Buchhave\altaffilmark{1}
Bertram Bitsch\altaffilmark{2},
Anders Johansen\altaffilmark{2},
David W. Latham\altaffilmark{3},
Martin Bizzarro\altaffilmark{4},
Allyson Bieryla\altaffilmark{3},
David M. Kipping\altaffilmark{5},
}

\altaffiltext{1}{DTU Space, National Space Institute, Technical University of Denmark, Elektrovej 328, DK-2800 Kgs. Lyngby, Denmark; \href{mailto:buchhave@space.dtu.dk}{buchhave@space.dtu.dk}}
\altaffiltext{2}{Lund Observatory, Lund University, Box 43, 221 00 Lund, Sweden}
\altaffiltext{3}{Harvard-Smithsonian Center for Astrophysics, 60 Garden Street, Cambridge, Massachusetts 02138, USA}
\altaffiltext{4}{Centre for Star and Planet Formation, Natural History Museum of Denmark, University of Copenhagen, Øster Voldgade 5-7, DK-1350 Copenhagen K, Denmark}
\altaffiltext{5}{Department of Astronomy, Columbia University, 550 W 120th Street, New York NY 10027, USA}

\begin{abstract}

Jupiter played an important role in determining the structure and configuration of the Solar System. Whereas hot-Jupiter type exoplanets preferentially form around metal-rich stars, the conditions required for the formation of planets with masses, orbits and eccentricities comparable to Jupiter (Jupiter analogues) are unknown. Using spectroscopic metallicities, we show that stars hosting Jupiter analogues have an average metallicity close to solar, in contrast to their hot-Jupiter and eccentric cool Jupiter counterparts, which orbit stars with super-solar metallicities. Furthermore, the eccentricities of Jupiter analogues increase with host star metallicity, suggesting that planet-planet scatterings producing highly eccentric cool Jupiters could be more common in metal-rich environments. To investigate a possible explanation for these metallicity trends, we compare the observations to numerical simulations, which indicate that metal-rich stars typically form multiple Jupiters, leading to planet-planet interactions and, hence, a prevalence of either eccentric cool Jupiters or hot-Jupiters with circularized orbits. Although the samples are small and exhibit variations in their metallicities, suggesting that numerous processes other than metallicity affect the formation of planetary systems, the data in hand suggests that Jupiter analogues and terrestrial-sized planets form around stars with average metallicities close to solar, whereas high metallicity systems preferentially host eccentric cool Jupiter or hot-Jupiters, indicating higher metallicity systems may not be favorable for the formation of planetary systems akin to the Solar System.

\end{abstract}

\section{Introduction}
\label{sec:intro}

Jupiter had a significant impact on the evolution of the Solar System, shaping its structure and configuration \citep{tsiganis_origin_2005} and possibly reducing the accretion rate of terrestrial planets by limiting inward transport of solids \citep{morbidelli_fossilized_2016,kooten_isotopic_2016}. Furthermore, gravitational perturbations from Jupiter were likely responsible for seeding the Earth with water-rich asteroids \citep{morbidelli_building_2012}. Recent observational advances in the exoplanet field have revealed scores of closely-packed multi-planet systems with short orbital periods and have shown that terrestrial-sized exoplanets are abundant \citep{borucki_kepler_2010,mayor_harps_2011}. However, it is still unclear if the configuration of the Solar System, with the inner region populated by terrestrial planets and the outer region by ice and gas giants, is common. 

Gas giant planets with masses, orbits and eccentricities analogous to Jupiter are difficult to detect due to their long orbital periods, but recent discoveries have increased the population of these Jupiter analogues enabling studies of the environment in which they formed. It is well-established that host-star metallicity, a proxy for the amount of heavy elements available to form planets in protoplanetary disks, is one of the driving factors determining the outcome of planet formation. The planet-metallicity correlation was first studied for hot-Jupiter type gas-giant planets \citep{santos_spectroscopic_2004,fischer_planet-metallicity_2005} and later for smaller sub-Neptune sized planets \citep{sousa_spectroscopic_2011,buchhave_abundance_2012,buchhave_three_2014}. However, the gas-giant studies were restricted to shorter period planets (e.g. periods $<$ 4 yr, \citep{fischer_planet-metallicity_2005}) and consequently the conditions promoting the formation of planetary systems containing Jupiter analogues are poorly understood. This information is critical to constrain planet formation theories and place the Solar System in the context of the multitude of different types of planetary systems that exist in the Galaxy. 

Here, we present a homogeneous analysis of host star metallicities for systems hosting planets with masses, orbits and eccentricities comparable to that of Jupiter, enabled by recent discoveries of these types of planets. Such data provide insights into the formation pathways of the largest and most massive planets that may significantly impact the architecture of their host planetary systems.

\section{Observations and Sample}
\label{sec:observations}

We define Jupiter analogues to be planets with approximately the mass of Jupiter ($0.3~\mjup < \mpl < 3.0~\mjup$, with $\mjup$ indicating the mass of Jupiter), and with low orbital eccentricities ($e < 0.25$) and receiving less than a quarter of the insolation of the Earth:
\begin{equation}
\seff = \frac{\teff}{5778~\kelvin} \big(\frac{\rstar}{\rsun} \big)^{2} \big(\frac{a}{1~\au} \big)^{-2} \frac{1}{\sqrt{1-e^2}} < 0.25~\searth
\end{equation}
with $\searth$ indicating units of Earth insolation. The low orbital eccentricity suggests that the systems are less likely to have undergone a period of strong dynamical instability. 

The insolation requirement would, in the Solar System, translate to an object having a semi-major axis $a > 2~\au$, which is beyond the snow line after the protoplanetary disk has evolved for a few million years \citep{bitsch_structure_2015}. Planetary growth rates are believed to increase outside of the water ice line due to the additional solid material in this region \citep{pollack_formation_1996} and the larger sizes of icy pebbles relative to silicate pebbles interior of the ice line \citep{morbidelli_great_2015}.

We used data from exoplanet.org\footnote{From exoplanets.org on 17 June, 2016} to identify currently known Jupiter analogue type exoplanets according to these criteria. We identified 20 Jupiter analogues and an additional 17 cool eccentric Jupiters with the same criteria except requiring that the cool eccentric Jupiter have an orbital eccentricity higher than 0.25. 

We collected a total of 1889 high-resolution spectra, including both new observations and publicly available archival spectra, from four different spectrographs of the host stars of 35 systems containing the 37 Jupiter analogues and eccentric cool Jupiters as well as a representative comparison sample of similar size of hot-Jupiter host stars. The hot-Jupiter sample consists of 28 hot-Jupiter host stars from the HATNet transit survey and from hot-Jupiter planets identified by the Kepler Mission that already have published metallicities derived using SPC \citep{buchhave_three_2014}. The hot-Jupiters were selected using the same mass restrictions as the Jupiter analogues and the additional requirement of a small semi-major axis ($a < 0.1~\au$).

The spectra originate from the following four spectrographs: the fiber-fed Tillinghast Reflector Echelle Spectrograph \citep[TRES;][]{furesz_design_2008} on the 1.5 m Tillinghast Reflector at the Fred Lawrence Whipple Observatory on Mt. Hopkins, Arizona; the High Accuracy Radial velocity Planet Searcher \citep[HARPS;][]{mayor_setting_2003} at the ESO La Silla 3.6 m telescope; the HIRES spectrograph \citep[HIRES;][]{vogt_hires:_1994} on the 10 m Keck I telescope at Mauna Kea, Hawaii and the FIber-fed Échelle Spectrograph \cite[FIES;][]{telting_fies:_2014} on the 2.5 m Nordic Optical Telescope (NOT) on La Palma, Spain. We gathered 21 new and 142 archival observations from TRES, 1689 public archival spectra from HARPS, 30 public archival spectra from HIRES and 7 archival spectra from FIES totalling 1889 spectra in all.

In order to minimize known biases associated with using different techniques to derive host star metallicities, we analyzed the gathered spectra of the host stars using the Stellar Parameter Classification tool \citep[SPC;][]{buchhave_abundance_2012,buchhave_three_2014}, deriving spectroscopic metallicities in a homogeneous and consistent manner for the entire sample of stars. SPC has been utilized in a large number of exoplanet publication and has been compared with other classification techniques by a number of independent research groups (e.g. \cite{huber_revised_2014, wang_calibration_2016,petigura_california-kepler_2017}).

We removed a total of 4 of the Jupiter analogues due to the reasons described in the following. SPC provides reliable stellar parameters for solar-type stars, but it has not been rigorously tested for evolved stars and cooler dwarf stars. We therefore restricted our analysis to solar-like dwarf stars ($\teff > 4500~\kelvin$ and $\logg > 4.0$) and thus removed two stars (HD114613 and HD11964) due to their low surface gravity ($\logg < 4.0$). Secondly, we were unable to obtain appropriate spectra of the two stars HD154857 and HIP57274 during the observing time available to us. The sample of hot-Jupiters also adheres to these restrictions on effective temperature and surface gravity.

We have compared our metallicities of the Jupiter analogues and the hot-Jupiters to the published values on exoplanet.org and we find an average difference of -0.01 dex and an RMS of the difference between our metallicities and the published metallicities of 0.07 dex, consistent with the formal uncertainties. Furthermore, we compared our metallicities to the SWEET-Cat database of stellar parameters\footnote{https://www.astro.up.pt/resources/sweet-cat} \citep{santos_sweet-cat:_2013} and found similar good agreement with an average difference of -0.03 dex and an RMS of the difference between our metallicities and the SWEET-Cat metallicities of 0.07 dex. Table \ref{tab:table1} contains the stellar parameters derived using SPC for all the planet hosting stars using in this article as well as values used in this manuscript from exoplanet.org at the time we downloaded them (17 June, 2016). Furthermore, we used previously published SPC metallicities for the Kepler hot-Jupiter planets \citep{buchhave_three_2014}.

To understand the formation history of the systems, including whether these have undergone dynamical planet-planet interactions that would result in large eccentricities, the planets are categorized using the eccentricity of the discovered planet with the maximum eccentricity in the system. However, when using the eccentricities of the individual planets, we find very similar results.

\section{Spectroscopic metallicities}
\label{sec:spectroscopicMetallicities}

\begin{figure*}
	\begin{center}
		\includegraphics[width=18cm]{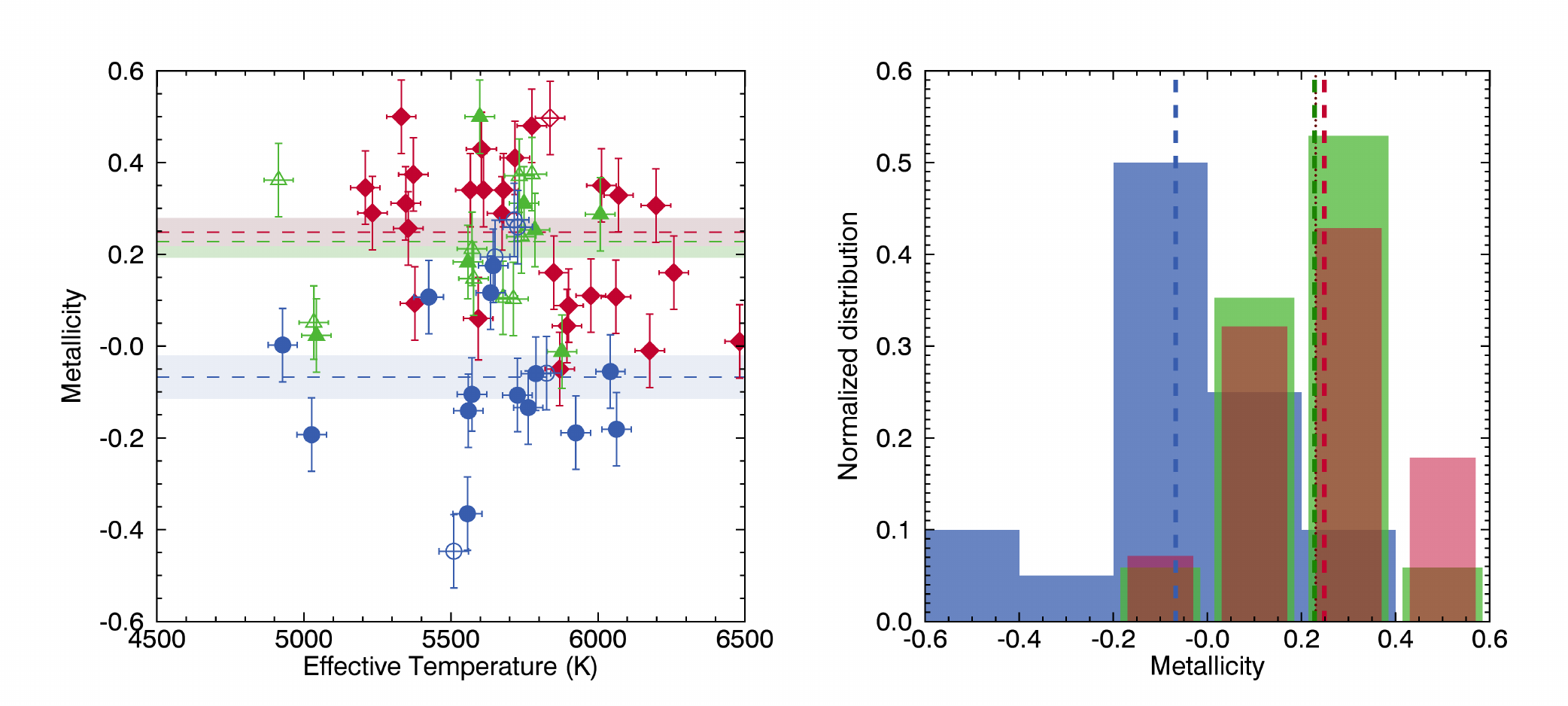}
		\caption{Left panel: Metallicities of Jupiter analogues (blue points), hot-Jupiters (red) and eccentric cool Jupiters (green) vs. their host star effective temperatures. The solid symbols represent systems with only one known planet, whereas the open symbols represent multi-planet systems with one or more additional planets.  The dashed lines indicate the average metallicity of the host stars and the shaded regions illustrate the standard error of the mean (Jupiter analogues: $-0.07 \pm 0.05~(\pm0.21)$, hot-Jupiters: $+0.25 \pm 0.03~(\pm 0.16)$ and eccentric cool Jupiters: $+0.23 \pm 0.04~(\pm 0.14)$, where the uncertainty is the standard error of the mean and the standard deviation is shown in parentheses). Right panel: Normalized histograms showing the metallicities of the host stars of the Jupiter analogues (blue), hot-Jupiters (red) and eccentric cool Jupiters (green). The widths of the red and green histogram columns have been reduced for visual clarity. The dashed lines show the average metallicities of the three populations. The dotted dark red line shows the average metallicity of the hot-Jupiter planets ($P < 10~\days$) from \cite{fischer_planet-metallicity_2005} ($\mh = +0.23 \pm 0.03$) discovered using the radial velocity technique, which is nearly identical to the average metallicity of the hot-Jupiter and eccentric cool Jupiter populations in thus study ($\mh = +0.25 \pm 0.03~(\pm 0.16)$ and $\mh = +0.23 \pm 0.04~(\pm 0.14)$, respectively). A two-sample Kolmogorov-Smirnov test reveals that the null hypothesis that the sample of stars orbited by Jupiter analogues and the sample of stars orbited by hot-Jupiters are draw from the same parent population can be rejected with 99.996\% confidence ($4.1\sigma$, $\pvalue = 4.26 \times 10^{-5}$). It is debated whether the migration of hot-Jupiters is due to planet-disk interactions \citep{baruteau_planet-disk_2014} (disk migration) or dynamical planet-planet interactions resulting in highly eccentric orbits that are subsequently circularized \citep{chatterjee_dynamical_2008}. Jupiter analogues are, on average, found around solar-metallicity host stars, in contrast to the hot-Jupiters, suggesting that the mechanism responsible for their proximity to the host stars must be more effective in high-metallicity environments.}
		\label{fig:fig1}
	\end{center}
\end{figure*}

In Figure \ref{fig:fig1} we show that the average host-star metallicity of Jupiter analogues is close to solar\footnote{SPC utilizes of all the absorption lines in the wavelength region 505 to 536 nm and we therefore denote the metallicities in this work as [m/H], representing a mix of metals assumed to be the same as the relative pattern of the abundances in the Sun.} ($\mh = -0.07 \pm 0.05~(\pm0.21)$), where the uncertainty is the standard error of the mean and the standard deviation is shown in parentheses), whereas the host stars of the hot-Jupiters and eccentric cool Jupiters are, on average, metal rich ($\mh = +0.25 \pm 0.03~(\pm 0.16)$ and $\mh = +0.23 \pm 0.04~(\pm 0.14)$, respectively). The average metallicity of the hot-Jupiter hosts is similar to the average metallicity of hot-Jupiter hosts from the RV studies and the Kepler Mission analyzed in previous studies (RV hot-Jupiters $\mh = +0.23 \pm 0.03$ \citep{fischer_planet-metallicity_2005} and Kepler gas-giants $\mh = +0.18 \pm 0.02$ \citep{buchhave_three_2014}). A two-sample Kolmogorov-Smirnov test reveals that the null hypothesis that the sample of stars orbited by Jupiter analogues and the sample of stars orbited by hot-Jupiters are drawn from the same parent population can be rejected with 99.996\% confidence ($4.1\sigma$, $\pvalue = 4.26 \times 10^{-5}$). Using the eccentricity of the individual planets as opposed to the maximum eccentricity in each system, we find that the confidence level of the K-S test is slightly lower at $3.9\sigma$. We also calculated the two independent sample t-test to determine the probability that two sample populations have significantly different means. For the Jupiter analogues and hot-Jupiters, we find that the null hypothesis can be rejected with 99.999\% confidence ($4.3\sigma$, $\pvalue = 1.44 \times 10^{-5}$) and with 99.997\% confidence ($4.2\sigma$, $\pvalue = 2.65 \times 10^{-5}$) when using the eccentricity of the individual planets.

\begin{figure*}
	\begin{center}
		\includegraphics[width=18cm]{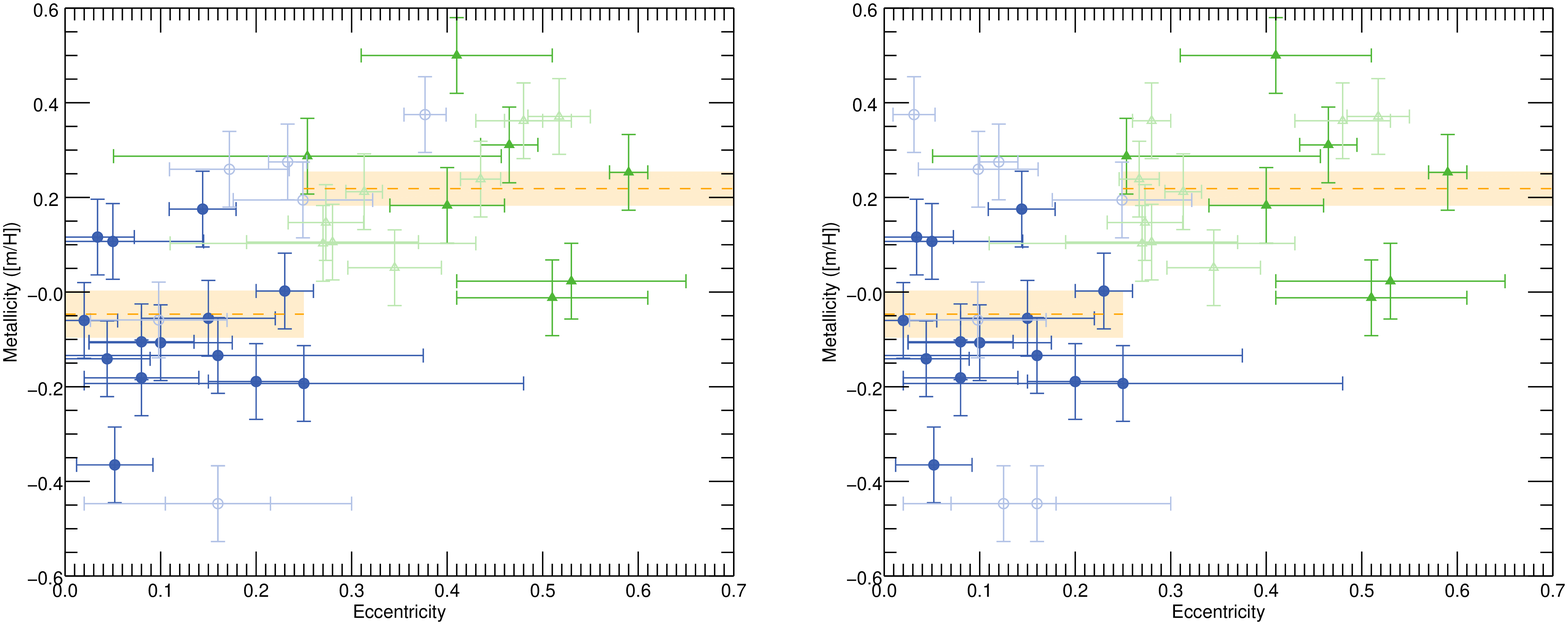}
		\caption{The eccentricity and metallicity of the Jupiter analogues (blue circles) and eccentric cool Jupiters (green triangles). The solid symbols represent systems with only one known planet, whereas the open symbols with lighter color represent multi-planet systems with one or more additional planets (the majority of which are Jupiter mass). In the left panel, the eccentricity reflects that of the largest eccentricity of the discovered planets in the system. In the right panel, the eccentricity assigned is the eccentricity of the individual planets. A two-sample K-S test shows that the null hypothesis that the sample of stars orbited by Jupiter analogues and the sample of stars orbited by cool eccentric Jupiters are drawn from the same parent population can be rejected with 99.96\% confidence ($3.6\sigma$, $\pvalue = 3.73 \times 10^{-4}$). Using the eccentricities of the individual planets rather than the maximum eccentricity of the system returns a comparable confidence level of 99.91\% ($3.3\sigma$, $\pvalue = 9.04 \times 10^{-4}$). To test whether a two-component model with two averages and with a dividing eccentricity as a free parameter is a better fit to the data than a simple average or a linear model, we fitted the data using a Markov Chain Monte Carlo (MCMC) approach to test three models, namely a simple average for the entire sample, a two-component model with two averages and with the dividing eccentricity as a free parameter and, lastly, a linear model. A two-component model with two averages and a fitted eccentricity divide (the orange dashed lines indicate the two averages and the shaded region indicates the standard error of the mean) is significantly better represented by the data than a model with a simple average ($\dbic = 120$, see Section \ref{sec:statisticalTests} for details).}
		\label{fig:fig2}
	\end{center}
\end{figure*}

Figure \ref{fig:fig2} shows the eccentricities of the Jupiter analogues plotted against their host star metallicities. The cool Jupiters with larger eccentricities are found orbiting host stars that are more metal rich, whereas Jupiter analogues with orbital eccentricities closer to circular in systems with only one discovered planet are predominantly found orbiting host stars with metallicities closer to that of the Sun. When including systems with multiple planets and assigning the eccentricities of the individual planets (right panel, all points), the planets appear to form around a wider range of metallicities.

The eccentricities of Jupiter analogues thus depend on host star metallicity and hence the initial metallicity of the protoplanetary disk, in agreement with previous findings for warm Jupiters (orbital distances of $0.1 - 1~\au$) \citep{dawson_giant_2013} and for short-period planet candidates from Kepler \citep{shabram_eccentricity_2016}, suggesting that planet-planet scatterings producing highly eccentric cool Jupiters are more common in metal rich environments. A two-sample K-S test shows that the null hypothesis that the sample of stars orbited by Jupiter analogues and the sample of stars orbited by cool eccentric Jupiters are drawn from the same parent population can be rejected with 99.96\% confidence ($3.6\sigma$, $\pvalue = 3.73 \times 10^{-4}$). Using the eccentricities of the individual planets rather than the maximum eccentricity of the system, the confidence level is 99.91\% ($3.3\sigma$, $\pvalue = 9.04 \times 10^{-4}$). The t-test null hypothesis can be rejected with 99.994\% confidence ($4.0\sigma$, $\pvalue = 6.47 \times 10{^-5}$) and with 99.987\% confidence ($3.8\sigma$, $\pvalue = 1.29 \times 10^{-4}$) when using the eccentricity of the individual planets. To test the robustness of our results to larger and smaller gas-giant masses, we expanded the mass range to $0.2~\mjup < \mpl < 13~\mjup$ which lead to similar results as above (see Section \ref{sec:statisticalTests} for details). Furthermore, bootstrap with replacement simulations indicate that the results are robust to the removal of individual data points (see Section \ref{sec:statisticalTests} for details).

Our results indicate that Jupiter analogues are found on average around stars with solar-metallicity in contrast to the hot-Jupiters that predominantly orbit metal-rich stars. Thus, the widely-accepted hot-Jupiter metallicity correlation does not seem to extend to longer period gas giants with low eccentricity. Hot-Jupiter exoplanets are thought to form beyond the ice line and subsequently migrate inwards close to their host stars. However, it is debated whether their migration is due to planet-disk interactions \citep[disk migration;][]{baruteau_planet-disk_2014} or dynamical planet-planet interactions resulting in highly eccentric orbits that are subsequently circularized \citep{chatterjee_dynamical_2008}. The results in Figure \ref{fig:fig1} suggest that the mechanism responsible for migration is more effective in high-metallicity regimes relative to environments with solar-like metallicities. Furthermore, Figure \ref{fig:fig2} shows that the eccentricity of Jupiter analogues is correlated with host star metallicity and thus the initial metallicity of the protoplanetary disc, suggesting that planet-planet scatterings producing highly eccentric Jupiter analogues is more common in metal rich environments. 

\section{Planet Formation Interpretation}
\label{sec:Interpretation}

These findings provide insights into the formation mechanisms responsible for gas giant planet formation and, as such, we performed numerical simulations of planet formation by pebble accretion to interpret our observations. We emphasize that this is one of many possible interpretations and should be viewed as a model that is consistent the observed metallicity trends. Recent work has shown that rapid formation of planetary cores within the lifetime of protoplanetary disks can occur by pebble accretion, that is, the accretion of centimetre- to meter-sized particles loosely bound to the gas onto planetesimals seeds \citep{lambrechts_rapid_2012,ormel_effect_2010,lambrechts_separating_2014,levison_growing_2015,bitsch_growth_2015}. Astronomical observations indicate that pebbles are abundant in the protoplanetary disks of dust and gas orbiting young stars \citep{brauer_survival_2007}. Moreover, pebble accretion allows for the efficient formation of planetary cores in disks with a wide range of metallicity as well as out to large orbital distances ($10-20~\au$) \citep{lambrechts_forming_2014,levison_growing_2015}. We note that our results could also be understood in the framework of planet formation models invoking collisional growth of planetesimals and planetary embryos.

\begin{figure*}
	\begin{center}
		\includegraphics[width=18cm]{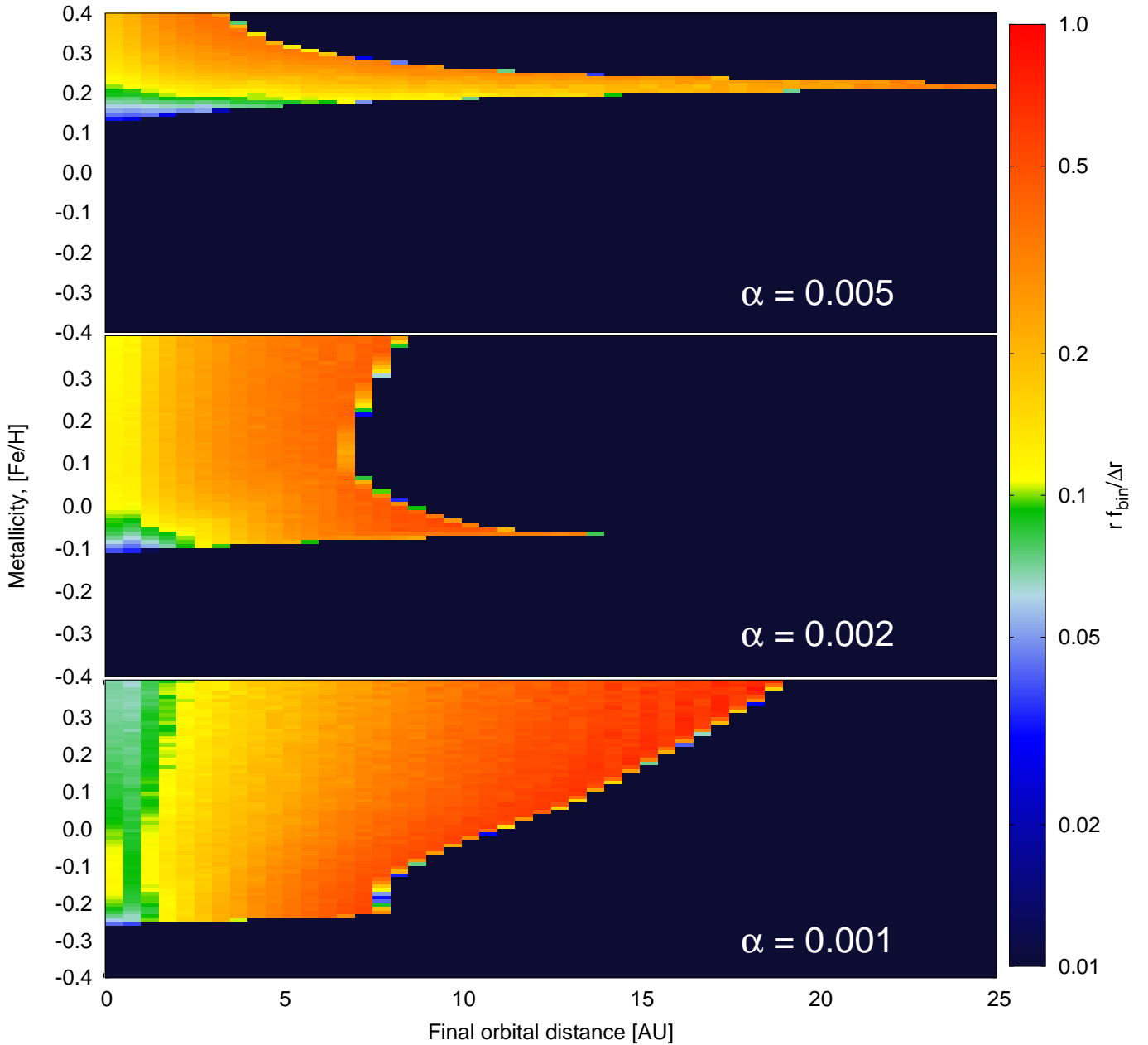}
		\caption{Fraction of planetary seeds that become gas-giant planets ($\mpl > 100~\mearth$) in planetary formation simulations, divided by the bin size and multiplied by the final orbital position to obtain the fraction per logarithmic orbital distance. This fraction is plotted as a function of final planetary position after migration and metallicity. The panels show results for three values of the dimensionless protoplanetary disk viscosity $\alpha$ (top: $\alpha=0.005$, middle: $\alpha=0.002$, bottom: $\alpha=0.001$). Each metallicity line contains 100,000 planetary seeds with random starting positions distributed in the range of 0.1-50 AU. The gas-giant fraction increases with increasing metallicity of the disc and with decreasing viscosity. At $\alpha$=0.002, approximately 20-25\% of all seeds form Jupiter analogues at solar metallicity ([Fe/H] = 0), but the gas-giant fraction rises sharply to above 50\% at slightly super solar values of the metallicity. Hence the formation of systems containing multiple cold gas giants is much more likely at elevated metallicities. Mutual scattering between these gas giants leads to ejections from the system and increases the eccentricity of the surviving planets; this can even lead to the formation of hot-Jupiter planets by tidal circularization at perihelion. This explains the trend for host stars of eccentric cool Jupiters and hot-Jupiters to have elevated metallicities. True Jupiter analogues, on the other hand, emerge in systems that form only a single gas giant, a more probable outcome at solar and sub-solar metallicity.}
		\label{fig:fig3}
	\end{center}
\end{figure*}

Our planet formation model is based on core formation via pebble and gas accretion, planet migration and disk evolution \citep{bitsch_growth_2015}. Planetary cores grow through the accretion of pebbles \citep{lambrechts_forming_2014} until they reach pebble isolation mass, which is where the planet carves a partial gap in the gas disc and pebble accretion stops \citep{lambrechts_separating_2014}. The planet then starts to accrete a gaseous envelope until $M_{\mathrm{env}} \sim M_{\mathrm{core}}$, which is when runaway gas accretion sets in. At the same time, the planet migrates through the disc, where we take the fully unsaturated torques into account \citep{paardekooper_numerical_2011}. The structure of the protoplanetary disc, the accretion rates for pebble accretion and the planetary migration rates depend strongly on the metallicity \citep{lambrechts_forming_2014,paardekooper_numerical_2011,bitsch_structure_2015}. Here, we use a simplified disk model consisting of two power laws, one for the viscously dominated and one for the irradiation dominated regime \citep{ida_radial_2016}. This allows us additionally to probe the dependency on the disk's viscosity. These different parameters (metallicity and viscosity) strongly influence the individual growth tracks (evolution of planetary mass as a function of semi-major axis) of the individual planets. We therefore investigate planet formation as a function of metallicity and viscosity (Figure \ref{fig:fig3}). For solar metallicity ([Fe/H]=0) we use a pebble-to-gas ratio of $Z_{peb}=1.35\%$.

In each metallicity bin ($\Delta$[Fe/H] = 0.01 dex), we simulate the growth tracks of 100,000 planets that have linearly randomly spaced starting locations in the interval 0.1 - 50 AU. The initial planetary mass is set to the pebble transition mass, where pebble accretion starts to be efficient from the whole Hill radius of the planet \citep{ida_radial_2016}. We assume that the disc is already 1.5 Myr old and has a total lifetime of 3 Myr, which is well within the limits of the measurements of lifetimes of protoplanetary discs \citep{mamajek_initial_2009}. In principle, shorter/longer disc lifetimes can change the giant planet formation frequency \citep{ndugu_planet_2018}, but for clarity we investigate here only the influence of a change in metallicity on the giant planet formation efficiency. This allows us to investigate the influence of metallicity on planet formation. In Figure \ref{fig:fig3}, we show the fraction of giant planets ($\mpl > 100~\mearth$) compared to the total number of planetary seeds in each metallicity bin as a function of their final orbital distance and the metallicity. Clearly, decreasing viscosity and increasing metallicity enhances planetary growth. In fact, for the highest metallicity investigated ([Fe/H] = 0.4) nearly 90\% of all planetary seeds grow to become gas giants. For the $\alpha$-viscosity parameters of 0.001 and 0.002, it seems that no giant planets would form outside of 10-15 AU at high metallicities. In order to form giant planets that end up at orbital positions larger than 10-15 AU, the planetary seed would need to form outside of 50 AU, which is larger than the interval we used for our simulations. We assume that the protoplanetary disc photoevaporates after 3 Myr. We do not take into account the possibility that discs around low-metallicity stars could have a shorter life-time to X-ray photoevaporation \citep{ercolano_metallicity_2010}, but adding such a dependency would simply work to amplify our key result that high-metallicity stars are more likely to form multiple giant planets.

Figure \ref{fig:fig3} shows the fraction of planetary seeds that become gas giants ($\mpl > 100~\mearth$) formed in planetary formation simulations including pebble accretion, gas accretion and planetary migration. The growth to giant planets is enhanced for lower viscosities, because of (i) the higher surface density for discs with constant accretion rates and (ii) because of the lower pebble scale height, which allows a faster accretion rate. The gas-giant fraction is shown as a function of the final orbital distance of the planet, the metallicity and the $\alpha$-viscosity parameter of the disk (varied in the three panels). Decreasing the viscosity allows for a more efficient growth of the planetary seeds, because the gas column density must be higher in low viscosity discs, in order to drive the observed mass accretion rate onto the star. Increasing the disc metallicity shortens the core accretion time-scale via pebble accretion, because the protoplanetary disk has more available pebbles to accrete \citep{bai_effect_2010}. Although planetary seeds in our simulations start at random locations between 0.1 and 50 AU, the final orbital distances of the gas giant planets are concentrated in the inner regions of the protoplanetary disk. This crowding of giant planets in the inner region increases with metallicity, as a larger fraction of the seeds grow to become gas giants. Therefore, high-metallicity systems are expected to contain multiple gas giants that can perturb each other gravitationally. These perturbations can lead to enhanced planet-planet scattering rates that will either eject one of the planets or leave it on an eccentric orbit and transfer the other planet to a much shorter orbital period \citep{lega_early_2013}, which has also been suggested by studies searching for outer gas-giant companions to known exoplanets \citep{bryan_statistics_2016}. Increased metallicity can also result in an enhancement of the planetesimal formation rate \citep{bai_effect_2010}. These leftover planetesimals can destabilize the giant planet configuration over longer time-scales \citep{morbidelli_chaotic_2005} and further increase the eccentricity distribution of the planetary systems. 

The enhanced planet-planet scattering observed in the simulations for higher metallicity systems due to the crowding of giant planets in the inner disk regions is in keeping with the observation that Jupiter analogues are found, on average, around stars of lower metallicity relative to the hot-Jupiter or eccentric cool Jupiter exoplanets. Eccentric cool Jupiters likely experienced dynamical encounters with other gas giants, resulting in the scattering of one of the planets. Thus, only a fraction of the gas-giant planets that originally formed around high metallicity stars remain Jupiter analogues. These observations suggest that both inward migration and planet-planet scatterings, resulting in hot-Jupiters and eccentric cool Jupiter exoplanets respectively, are more common in metal-rich systems. Therefore, the initial metal content of protoplanetary disks appears to regulate the structure and architecture of planetary systems. 

Another possible interpretation is that the metallicity trends could be an imprint of disk dispersal, where disk lifetimes are shorter at lower metallicities \citep{yasui_lifetime_2009,ercolano_metallicity_2010}. The short lifetime could, on average, yield fewer multiple giant planets, and these single giant planets could migrate inward and their final location would be set by inside-out photoevaporation which produces an increase of single giant planets just beyond the critical radius $~1~\au$ \citep{alexander_deserts_2012}, corresponding to the cool circular orbit Jupiters in this analysis. The timescale for dispersing the high-metallicity disks is longer and there is therefore more time to form multiple large planets whose final locations (and eccentricities) depend on the interaction with other planets, both while migrating in the disk as well as post disk dispersal.

\section{Additional Tests and Discussion}
\label{sec:statisticalTests}

In order to test whether excluding the two stars with lower surface gravity $\logg < 4.0$ affected our results, we lowered our surface gravity threshold to 3.8, which resulted in HD114613 and HD11964 being included in the analysis. The K-S test described in Section \ref{sec:spectroscopicMetallicities} decreased slightly in significance but remained statistically significant. Subsequently, we used stellar parameters from SWEET-Cat \citep{santos_sweet-cat:_2013} in order to include HD154857 and HIP57274 in our analysis. In this case, the significance of the results in the main paper increased slightly.

The Jupiter analogues and cool eccentric Jupiter planet discoveries originate from a number of RV survey programs designed to discover and measure the mass of exoplanets using precise radial velocity measurements. All of the surveys target solar-type stars that are bright and in proximity to our solar system, and some of the surveys, like the HARPS and CORALIE surveys, are volume-limited surveys, thus targeting all solar-type stars within a given distance. For example, the HARPS volume-limited sample \citep{naef_harps_2010} contains 850 F8 to M0 dwarfs within 57.5 pc. The sample of hot-Jupiter planets in this paper were discovered using the transit method and subsequently followed up with radial velocities to measure their masses. Both the RV surveys and the transit surveys suffer from inherit detection biases: larger and more massive planets in short orbital periods are more easily detected. Furthermore, there are human biases, such as prioritizing targets that exhibit variability early on because they might yield publishable results. However, the vast majority of the surveys (both RV and transit) have no significant inherent bias towards metallicity, since the samples from the majority of the surveys were selected as volume limited samples or targeted bright solar-like stars.

Detection biases affect occurrence rate calculations and therefore, completeness studies must be undertaken to account for these effects in order to draw robust conclusions. However, this study pertains to the host star metallicities of exoplanets belonging to different categories of planets (Jupiter analogues, eccentric cool Jupiter and hot-Jupiters) and does not address occurrence rate questions. When surveys are biased towards discovering a particular type of planet (like hot-Jupiters), the detected number of planets of this particular type of planets is affected (and hence the occurrence rate, if the bias is not dealt with), but the average metallicity of the host stars of a particular type of planet (e.g. hot Jupiters) are not affected by this bias, unless there is a very strong inherent bias towards observing only a particular host star metallicity, which is not the case.

The surveyed stars in the Kepler Mission were not selected based on host star metallicity and neither were the surveyed stars in the ground-based transit surveys nor the RV surveys. It is also important to note that even if there was a bias towards, for example, more readily detecting planets around metal rich stars for some reason, the results in this paper would be even more significant, since it would disfavor the detection of Jupiter analogues, which form around stars with lower average metallicities. 

Furthermore, we note that the hot-Jupiter sample in this survey, which is based on planets detected using the transit method, have an average metallicities nearly identical to the hot-Jupiters in the \cite{fischer_planet-metallicity_2005} paper discovered using the radial velocity method (average metallicity of hot-Jupiters from \cite{fischer_planet-metallicity_2005} $\mh = +0.23 \pm 0.03$ and average metallicity of hot-Jupiters in this paper $\mh = +0.25 \pm 0.03$). As such, biases inherit to the detection methods should not affect the result and conclusions in this paper, since we are examining the average metallicities of the host stars of planets of different types.

To test if large uncertainties on the eccentricities affect our results, we imposed a maximum eccentricity uncertainty of 0.2, which resulted in the removal of two systems. This resulted in a change in the confidence level of the K-S test from 99.996\% ($4.1\sigma$, $\pvalue = 4.26 \times 10^{-5}$) to 99.97\% ($3.7\sigma$, $\pvalue = 2.57 \times 10^{-4}$).

In order to ensure the robustness of the results, we performed a bootstrap with replacement analysis. We randomly drew new samples from the original samples (of the same size as the original samples) and allowed planets to be drawn more than once (“with replacement”) and repeated the test $10^6$ times. We then performed the same K-S test for each of the draws asking with what confidence the null hypothesis  that the newly drawn sample of stars orbited by Jupiter analogues and the sample of stars orbited by hot-Jupiters were drawn from the same parent population could be rejected. We found that the samples were not drawn from the same parent population with $3.98^{+0.60}_{-0.67} \sigma$ confidence, in line with the conclusions in Section \ref{sec:spectroscopicMetallicities}.

To further test the robustness of our results to different mass definitions of Jupiter analogues, we expanded our mass range to included planets with masses between $0.2~\mjup < \mplsini < 13~\mjup$. Using exoplanet.org, we found 2 planets with masses between $0.2~\mjup$ and $0.3~\mjup$ and 20 planets with masses between $3.0~\mjup$ and $13.0~\mjup$. To measure the host star metallicities of these stars, we gathered 9 new observations from TRES and 736 public archival spectra from HARPS, which yielded reliable host star metallicities for 17 of the 22 planets. Using the full mass range ($0.2~\mjup < \mplsini < 13.0~\mjup$), we find average metallicities of the Jupiter analogues (26 planets), the hot-Jupiters (32 planets) and the cool eccentric Jupiters (28 planets) of $\mh = -0.04 \pm 0.05$, $\mh = +0.24 \pm 0.03$ and $\mh = +0.17 \pm 0.03$, respectively. The two-sample K-S test shows that the null hypothesis that the sample of stars orbited by Jupiter analogues and the sample of stars orbited by hot-Jupiters are drawn from the same parent population can be rejected with 99.996\% confidence ($4.1\sigma$, $\pvalue = 3.92 \times 10^{-5}$). Furthermore, a two-sample K-S test shows that the null hypothesis that the sample of stars orbited by Jupiter analogues and the sample of stars orbited by cool eccentric Jupiters are drawn from the same parent population can be rejected with 99.90\% confidence ($3.3\sigma$, $\pvalue = 9.74 \times 10^{-4}$).

\begin{figure*}
	\begin{center}
		\includegraphics[width=18cm]{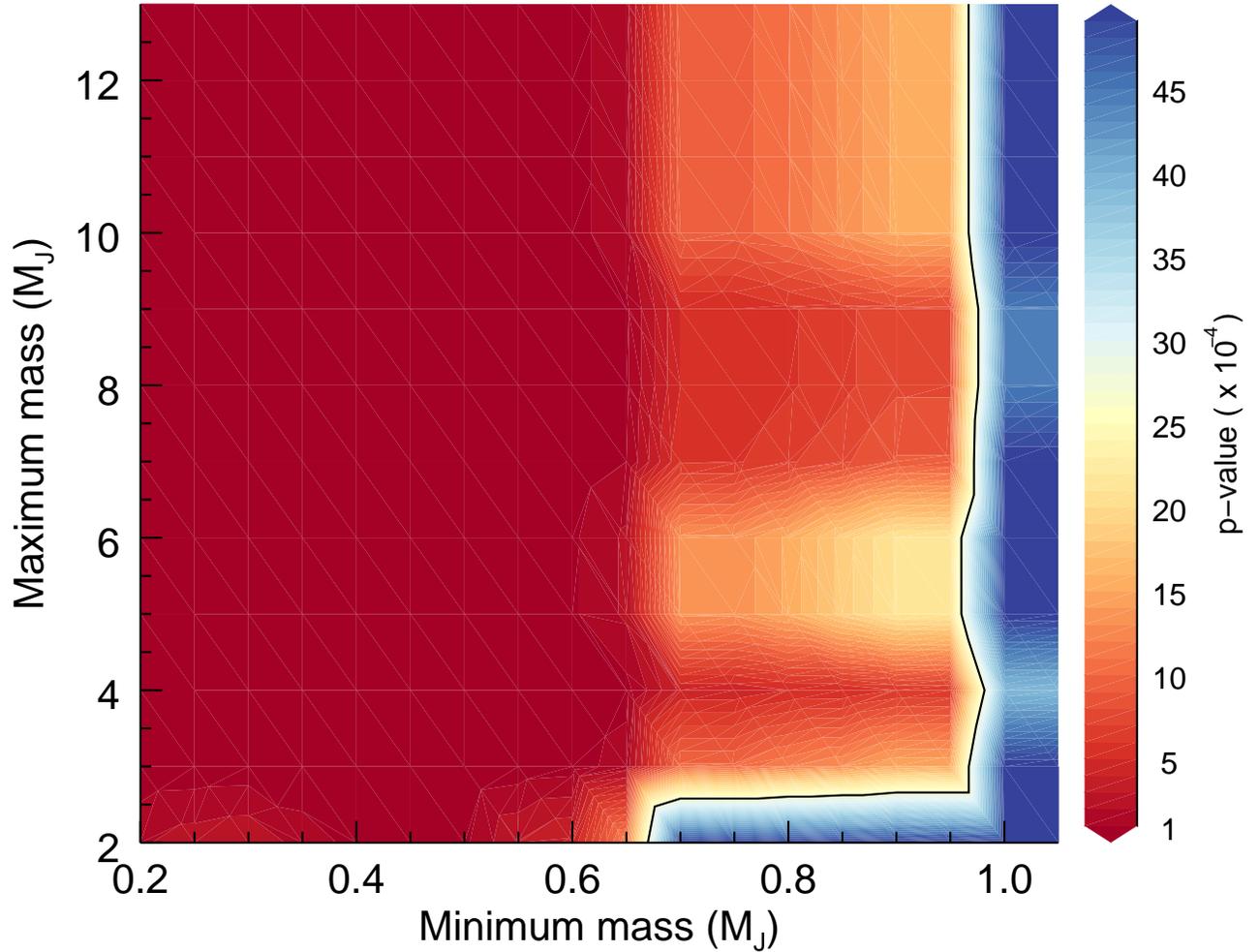}
		\caption{We performed K-S tests on the Jupiter analogues and the hot-Jupiters, testing various mass ranges. We computed the p-value from the K-S test in a grid varying the minimum mass from $0.2~\mjup$ to $1.1~\mjup$ in $0.05~\mjup$ steps and the maximum mass from $2~\mjup$ to $13~\mjup$ in $1~\mjup$ steps. The resulting p-values are shown as color coded contours with red having a high confidence level and blue a low confidence level. The red and orange colors encapsulated by the solid black line represent p-values corresponding to a confidence level over $3\sigma$.}
		\label{fig:fig4}
	\end{center}
\end{figure*}

We then performed K-S tests on the Jupiter analogues and the hot-Jupiters, testing various mass range combinations. We computed the $\pvalue$ from the K-S test in a grid varying the minimum mass from $0.2~\mjup$ to $1.1~\mjup$ in $0.05~\mjup$ steps and the maximum mass from $2~\mjup$ to $13~\mjup$ in $1~\mjup$ steps. The resulting $\pvalue$s can be seen in Figure \ref{fig:fig4}, where the red and orange colors encapsulated by the solid black line represent p-values corresponding to a confidence level over $3\sigma$.

To explore whether a two-component model with two averages and with a dividing eccentricity as a free parameter is a better fit to the data than a simple average or a linear model, we fitted the data using three different models using a Markov Chain Monte Carlo (MCMC) approach. We tested three models: a simple average for the entire sample, a two-component model with two averages and with the dividing eccentricity as a free parameter and, lastly, a linear model. We derived the maximum likelihood from the posteriors and calculated the Bayesian Information Criterion (BIC) for each model. 

We found that the two-component model significantly better represented the data than the model with a simple average ($\dbic = 120$). The linear model was also significantly preferred over the model with a simple average ($\dbic = 106$). The two-component model was thus statistically significantly favored over both the other models. Furthermore, in order to account for the uncertainties in the eccentricities, we created 5000 new sets of data by perturbing the eccentricities within their uncertainties. For each of the 5000 sets of data, we carried out the test described above and determined the BIC values from the posteriors of the distributions. We found similar results as above: the two-component model was significantly favored over the model with a simple average ($\dbic = 95.3_{-30.6}^{+23.8}$) and the linear model was similarly favored over the simple average ($\dbic =80.1_{-31.3}^{+25.8}$). And finally, as before, the two-component was on average also significantly preferred over the linear model, although with large uncertainties ($\dbic = 15.9_{-43.8}^{+35.1}$). These results, along with the results from the K-S test and the statistically significant difference in the average metallicities when considering the standard error of the mean of the two samples, shows that the results are indeed robust.

\begin{figure*}
	\begin{center}
		\includegraphics[width=18cm]{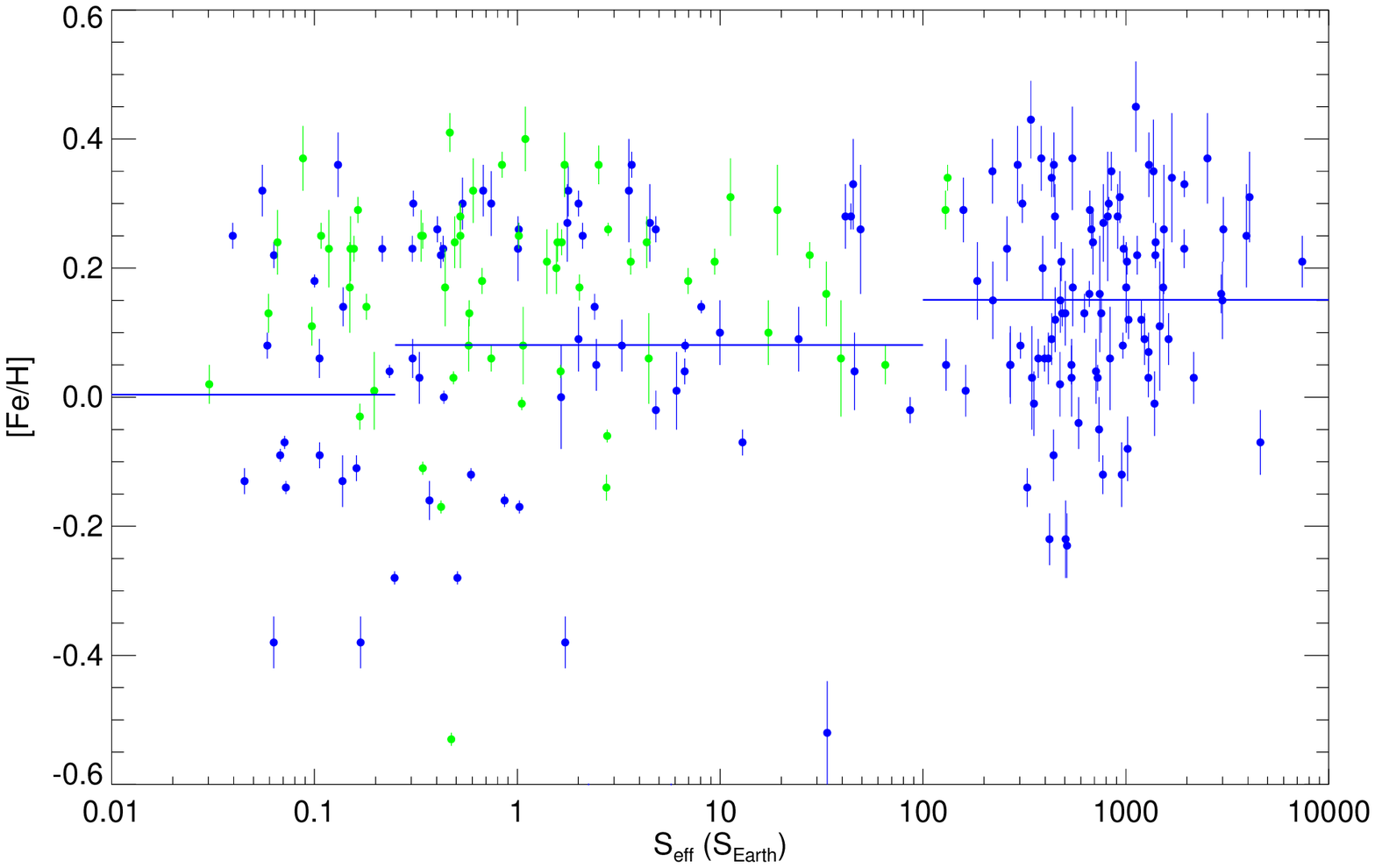}
		\caption{The metallicity of the host stars of planets with masses between $0.3~\mjup < \mpl < 3.0~\mjup$ from exoplanets.org as a function of the insolation received in units of Earth insolation. The metallicities are from the SWEET-Cat catalog using only stars that were flagged as having been homogeneously analyzed using the same method. Blue points signify stars with eccentricities e < 0.25 and green points are stars orbited by planets with $e > 0.25$. The horizontal blue lines indicate the average metallicity of stars with planets with low eccentricity receiving insolation of $0.01~\searth < \seff < 0.25~\searth$, $0.25~\searth < \seff < 10~\searth$ and $10~\searth < \seff < 10000~\searth$. The intermediate “warm” Jupiters have an average metallicity which is in between the Jupiter analogues and the hot-Jupiter type planets.  }
		\label{fig:fig5}
	\end{center}
\end{figure*}

In this work, we require that the Jupiter analogue planets receive less than a quarter of the insolation of the Earth ($\seff < 0.25~\searth$). To investigate the influence of this requirement on our results, we performed tests where we included planets receiving a larger amount of insolation by using literature values originating from SWEET-Cat \citep{santos_sweet-cat:_2013}, which is a collection of stellar parameters for planet hosting stars with the goal of providing a homogeneous table of stellar parameters when possible. In Figure \ref{fig:fig5}, we show the metallicities of stars hosing planets with masses between $0.3~\mjup < \mpl < 3.0~\mjup$ as a function of the insolation they receive. The average metallicities of the stars in receiving an insolation of (cool) $0.01~\searth < \seff < 0.25~\searth$, (warm) $0.25~\searth < \seff < 10~\searth$ and (hot) $10~\searth < \seff < 10000~\searth$ is (cool) $0.00 \pm 0.05~(\pm0.22)$, (warm) $+0.08 \pm 0.03~(\pm0.24)$, (hot) $+0.15 \pm 0.02~(\pm0.15)$, where the uncertainty is given as the standard error of the mean and the standard deviation is in parentheses. The cool non-eccentric Jupiter analogs and the hot-Jupiters show the same pattern as seen in Section \ref{sec:spectroscopicMetallicities}, although the hot-Jupiters are slightly less metal rich. The warm non-eccentric Jupiters appear to have an average metallicity intermediate to the cool and hot Jupiters. We note that when using period or semi-major axis as a proxy rather than insolation, the hot-Jupiters orbit more metal rich stars, while the differences between the medium and long period planets (or planets with a medium or large semi-major axis) is diminished, indicating the importance of using insolation to investigate the metallicity dependence. As such, these results seem compatible with recent results from the LAMOST survey, where \cite{mulders_super-solar_2016} find that the average metallicities of stars hosing large planets (larger than Neptune) appear to be consistently metal-rich out to a period of a few hundred days (see their Figure 8, bottom panel). However, the number of targets is small and although not statistically significant, it appears that the close-in hot-Jupiters orbit more metal rich host stars with a slight decrease in metallicity at longer orbital periods, which is consistent with our findings when examining planets from exoplanets.org and SWEET-Cat.

We note that the average orbital period of the Jupiter analogues is 2348 days whereas Jupiter has a longer orbital period of 4332 days. We also caution that the samples of Jupiter analogues and eccentric cool Jupiter in this analysis are small since such planets are difficult to detect with current techniques.

The S/N ratio of all the spectra used is adequate to derive precise stellar parameters using SPC. As such, the uncertainties in the stellar parameters reflect the systematics in the stellar models and not the photon noise. Thick disk stars have a wide variation in their metallicities, but their metallicity distribution peaks at [Fe/H] ~ -0.7. The stars in our sample are thus consistent with being thin disk stars. As such, we conclude that neither S/N ratio nor galactic chemical evolution could be the root cause of the average metallicity difference we observe.

\section{Conclusion}
\label{sec:conclusion}

Although a number of factors and initial conditions influence the outcome of planet formation, metallicity appears to be one of the key parameters determining which type of planets are formed and their final configuration. 
In this paper, we show, using homogeneously derived spectroscopic metallicities, that stars hosting Jupiter analogues have an average metallicity close to solar, in contrast to their hot-Jupiter and eccentric cool Jupiter counterparts, which orbit stars with super-solar metallicities. And warm Jupiters seem to orbit stars with an average metallicity intermediate to that of the hot and cool Jupiters. We also find that the eccentricities of Jupiter analogues increase with host star metallicity, suggesting that planet-planet scatterings producing highly eccentric cool Jupiters could be more common in metal-rich environments. Conversely, very metal poor systems will lack the required amount of material to form planets at all.

This suggests the existence of an intermediate metallicity regime where the formation of terrestrial planets and of a single gas giant akin to Jupiter is more likely to occur. The data at hand indicate that, on average, Jupiter analogues (this paper) and terrestrial size planets \citep{buchhave_abundance_2012,buchhave_three_2014,mulders_super-solar_2016} that are not in very short period orbits form around stars with lower metallicities that are, on average, close to solar.

A direct consequence of the formation of planetary cores (with at least 10-20 Earth masses) is their feedback onto the disc structure. The planet opens up a partial gap and generates a pressure bump outside of its orbit, which is an efficient mechanism for limiting the inward flux of pebbles \citep{lambrechts_separating_2014}. Thus, the amount of solids available to fuel planetary growth sunward of the gap may be significantly reduced, resulting in much slower growth rates in the inner disk \citep{morbidelli_great_2015}. If correct, this could imply that systems analogous to the Solar System, with small, rocky and potentially habitable planets might be common around stars hosting a Jupiter analogue.

\begin{acknowledgements}
MB acknowledges financial support from the Danish National Research Foundation (\#DNRF97) and from the European Research Council (ERC Consolidator grant agreement 616027-STARDUST2ASTEROIDS). LAB acknowledges financial support from the Villum Foundation YIP Program and the Carlsberg Foundation Distinguished Associate Professor Fellowship.
\end{acknowledgements}


\clearpage

\begin{deluxetable*}{lccccccccccc}
	\tabletypesize{\tiny} 
	\tablecaption{Parameters for Jupiter analogs, eccentric cool Jupiters and hot-Jupiters. \label{tab:table1}}
	\tablewidth{0pt}
	\tablecolumns{12}
	\tablehead{
		\colhead{Star} &
		\colhead{pl} &
		\colhead{Teff} &
		\colhead{$\mh$} &
		\colhead{$\logg$} &
		\colhead{ncomp} &
		\colhead{ecc} &
		\colhead{Max ecc} &
		\colhead{$\msini$} &
		\colhead{Period} &
		\colhead{a} &
		\colhead{$\rstar$} 
	}
		\startdata  
\multicolumn{12}{l}{\textbf{Jupiter analogs}} \\
47UMa  &       c & $5825 \pm 50$ & $-0.06 \pm 0.08$ & $4.26 \pm 0.10$ & 2 & $0.098 \pm 0.071$ & $0.098$ & $0.546 \pm 0.071$ & $2391.0 \pm  93.5$ & $3.572 \pm 0.111$ & $1.119 \pm 0.038$ \\
HD114386  &    b & $4928 \pm 50$ & $ 0.00 \pm 0.08$ & $4.62 \pm 0.10$ & 1 & $0.230 \pm 0.030$ & $0.230$ & $1.364 \pm 0.079$ & $ 937.7 \pm  15.6$ & $1.727 \pm 0.035$ & $0.637 \pm 0.029$ \\
HD117207  &    b & $5644 \pm 50$ & $ 0.18 \pm 0.08$ & $4.27 \pm 0.10$ & 1 & $0.144 \pm 0.035$ & $0.144$ & $1.819 \pm 0.089$ & $2597.0 \pm  41.0$ & $3.738 \pm 0.074$ & $0.952 \pm 0.034$ \\
HD12661  &     c & $5776 \pm 50$ & $ 0.38 \pm 0.08$ & $4.42 \pm 0.10$ & 2 & $0.031 \pm 0.022$ & $0.377$ & $1.949 \pm 0.092$ & $1707.9 \pm  13.9$ & $2.919 \pm 0.064$ & $1.068 \pm 0.037$ \\
HD134987  &    c & $5715 \pm 50$ & $ 0.27 \pm 0.08$ & $4.23 \pm 0.10$ & 2 & $0.120 \pm 0.020$ & $0.233$ & $0.805 \pm 0.046$ & $5000.0 \pm 400.0$ & $5.825 \pm 0.331$ & $1.166 \pm 0.039$ \\
HD13931  &     b & $5789 \pm 50$ & $-0.06 \pm 0.08$ & $4.21 \pm 0.10$ & 1 & $0.020 \pm 0.035$ & $0.020$ & $1.881 \pm 0.142$ & $4218.0 \pm 388.0$ & $5.149 \pm 0.327$ & $1.224 \pm 0.042$ \\
HD154345  &    b & $5559 \pm 50$ & $-0.14 \pm 0.08$ & $4.59 \pm 0.10$ & 1 & $0.044 \pm 0.045$ & $0.044$ & $0.957 \pm 0.061$ & $3341.6 \pm  92.7$ & $4.214 \pm 0.105$ & $1.000 \pm 0.050$ \\
HD164922  &    b & $5424 \pm 50$ & $ 0.11 \pm 0.08$ & $4.48 \pm 0.10$ & 1 & $0.050 \pm 0.095$ & $0.050$ & $0.358 \pm 0.060$ & $1155.0 \pm  23.0$ & $2.101 \pm 0.045$ & $0.899 \pm 0.033$ \\
HD222155  &    b & $5763 \pm 50$ & $-0.13 \pm 0.08$ & $4.10 \pm 0.10$ & 1 & $0.160 \pm 0.215$ & $0.160$ & $2.026 \pm 0.500$ & $3999.0 \pm 505.0$ & $5.139 \pm 0.464$ & $1.670 \pm 0.070$ \\
HD25171  &     b & $6063 \pm 50$ & $-0.18 \pm 0.08$ & $4.17 \pm 0.10$ & 1 & $0.080 \pm 0.060$ & $0.080$ & $0.956 \pm 0.234$ & $1845.0 \pm 167.0$ & $3.031 \pm 0.190$ & $1.069 \pm 0.041$ \\
HD290327  &    b & $5571 \pm 50$ & $-0.11 \pm 0.08$ & $4.42 \pm 0.10$ & 1 & $0.080 \pm 0.055$ & $0.080$ & $2.548 \pm 0.206$ & $2443.0 \pm 161.0$ & $3.431 \pm 0.161$ & $0.997 \pm 0.033$ \\
HD37124  &     c & $5510 \pm 50$ & $-0.45 \pm 0.08$ & $4.35 \pm 0.10$ & 3 & $0.125 \pm 0.055$ & $0.160$ & $0.648 \pm 0.055$ & $ 885.5 \pm   5.1$ & $1.710 \pm 0.029$ & $0.772 \pm 0.030$ \\
HD37124  &     d & $5510 \pm 50$ & $-0.45 \pm 0.08$ & $4.35 \pm 0.10$ & 3 & $0.160 \pm 0.140$ & $0.160$ & $0.687 \pm 0.075$ & $1862.0 \pm  38.0$ & $2.807 \pm 0.060$ & $0.772 \pm 0.030$ \\
HD4208  &      b & $5556 \pm 50$ & $-0.37 \pm 0.08$ & $4.24 \pm 0.10$ & 1 & $0.052 \pm 0.040$ & $0.052$ & $0.807 \pm 0.041$ & $ 828.0 \pm   8.1$ & $1.654 \pm 0.030$ & $0.877 \pm 0.032$ \\
HD47186  &     c & $5650 \pm 50$ & $ 0.19 \pm 0.08$ & $4.29 \pm 0.10$ & 2 & $0.249 \pm 0.073$ & $0.249$ & $0.348 \pm 0.076$ & $1353.6 \pm  57.1$ & $2.387 \pm 0.078$ & $1.131 \pm 0.037$ \\
HD6718  &      b & $5726 \pm 50$ & $-0.11 \pm 0.08$ & $4.37 \pm 0.10$ & 1 & $0.100 \pm 0.075$ & $0.100$ & $1.559 \pm 0.117$ & $2496.0 \pm 176.0$ & $3.554 \pm 0.177$ & $0.957 \pm 0.031$ \\
HD70642  &     b & $5635 \pm 50$ & $ 0.12 \pm 0.08$ & $4.27 \pm 0.10$ & 1 & $0.034 \pm 0.039$ & $0.034$ & $1.909 \pm 0.104$ & $2068.0 \pm  39.0$ & $3.181 \pm 0.066$ & $1.031 \pm 0.036$ \\
HD8535  &      b & $6042 \pm 50$ & $-0.06 \pm 0.08$ & $4.18 \pm 0.10$ & 1 & $0.150 \pm 0.070$ & $0.150$ & $0.682 \pm 0.052$ & $1313.0 \pm  28.0$ & $2.445 \pm 0.054$ & $1.044 \pm 0.036$ \\
HD89307  &     b & $5925 \pm 50$ & $-0.19 \pm 0.08$ & $4.33 \pm 0.10$ & 1 & $0.200 \pm 0.050$ & $0.200$ & $1.791 \pm 0.150$ & $2166.0 \pm  38.0$ & $3.266 \pm 0.067$ & $1.151 \pm 0.039$ \\
epsilonEri  &  b & $5026 \pm 50$ & $-0.19 \pm 0.08$ & $4.39 \pm 0.10$ & 1 & $0.250 \pm 0.230$ & $0.250$ & $1.054 \pm 0.188$ & $2500.0 \pm 350.0$ & $3.376 \pm 0.322$ & $0.740 \pm 0.010$ \\
muAra  &       c & $5727 \pm 50$ & $ 0.26 \pm 0.08$ & $4.20 \pm 0.10$ & 4 & $0.099 \pm 0.063$ & $0.172$ & $1.889 \pm 0.223$ & $4205.8 \pm 458.9$ & $5.341 \pm 0.402$ & $1.250 \pm 0.042$ \\
\multicolumn{12}{l}{\textbf{Jupiter analogs with expanded mass limits}} \\
55Cnc  &       d & $5382 \pm 50$ & $ 0.38 \pm 0.08$ & $4.44 \pm 0.10$ & 5 & $0.020 \pm 0.008$ & $0.320$ & $3.545 \pm 0.122$ & $4909.0 \pm  30.0$ & $5.475 \pm 0.094$ & $0.943 \pm 0.010$ \\
HD10180  &     h & $5806 \pm 50$ & $-0.01 \pm 0.08$ & $4.23 \pm 0.10$ & 6 & $0.151 \pm 0.072$ & $0.151$ & $0.206 \pm 0.016$ & $2248.0 \pm 104.0$ & $3.425 \pm 0.120$ & $1.109 \pm 0.036$ \\
HD111232  &    b & $5521 \pm 50$ & $-0.36 \pm 0.08$ & $4.44 \pm 0.10$ & 1 & $0.200 \pm 0.010$ & $0.200$ & $6.842 \pm 0.251$ & $1143.0 \pm  14.0$ & $1.975 \pm 0.037$ & $0.875 \pm 0.042$ \\
HD128311  &    c & $5033 \pm 50$ & $ 0.05 \pm 0.08$ & $4.66 \pm 0.10$ & 2 & $0.230 \pm 0.058$ & $0.345$ & $3.248 \pm 0.159$ & $ 923.8 \pm   5.3$ & $1.745 \pm 0.030$ & $0.583 \pm 0.028$ \\
HD183263  &    c & $5853 \pm 50$ & $ 0.20 \pm 0.08$ & $4.25 \pm 0.10$ & 2 & $0.239 \pm 0.064$ & $0.357$ & $3.476 \pm 0.309$ & $3066.0 \pm 110.0$ & $4.295 \pm 0.125$ & $1.117 \pm 0.038$ \\
HD204313  &    b & $5677 \pm 50$ & $ 0.11 \pm 0.08$ & $4.22 \pm 0.10$ & 2 & $0.230 \pm 0.040$ & $0.280$ & $3.501 \pm 0.221$ & $1920.1 \pm  25.0$ & $3.071 \pm 0.058$ & $1.126 \pm 0.038$ \\
HD24040  &     b & $5741 \pm 50$ & $ 0.09 \pm 0.08$ & $4.20 \pm 0.10$ & 1 & $0.040 \pm 0.065$ & $0.040$ & $4.022 \pm 0.326$ & $3668.0 \pm 170.0$ & $4.924 \pm 0.206$ & $1.154 \pm 0.039$ \\
HD30177  &     b & $5642 \pm 50$ & $ 0.42 \pm 0.08$ & $4.32 \pm 0.10$ & 1 & $0.193 \pm 0.025$ & $0.193$ & $9.688 \pm 0.544$ & $2770.0 \pm 100.0$ & $3.808 \pm 0.134$ & $1.212 \pm 0.041$ \\
HD37605  &     c & $5477 \pm 50$ & $ 0.31 \pm 0.08$ & $4.50 \pm 0.10$ & 2 & $0.013 \pm 0.014$ & $0.677$ & $3.366 \pm 1.124$ & $2720.0 \pm  57.0$ & $3.818 \pm 0.638$ & $0.917 \pm 0.030$ \\
HD72659  &     b & $5893 \pm 50$ & $-0.07 \pm 0.08$ & $4.16 \pm 0.10$ & 1 & $0.220 \pm 0.030$ & $0.220$ & $3.174 \pm 0.148$ & $3658.0 \pm  32.0$ & $4.754 \pm 0.084$ & $1.343 \pm 0.044$ \\
HD95872  &     b & $5361 \pm 50$ & $ 0.31 \pm 0.08$ & $4.56 \pm 0.10$ & 1 & $0.060 \pm 0.040$ & $0.060$ & $4.594 \pm 0.352$ & $4375.0 \pm 169.0$ & $5.154 \pm 0.158$ & $0.999 \pm 0.048$ \\
\multicolumn{12}{l}{\textbf{Eccentric cool Jupiters}} \\
HD108874  &    c & $5577 \pm 50$ & $ 0.15 \pm 0.08$ & $4.32 \pm 0.10$ & 2 & $0.273 \pm 0.040$ & $0.273$ & $1.028 \pm 0.054$ & $1680.4 \pm  23.9$ & $2.720 \pm 0.052$ & $1.121 \pm 0.038$ \\
HD126614A &    b & $5598 \pm 50$ & $ 0.50 \pm 0.08$ & $4.28 \pm 0.10$ & 1 & $0.410 \pm 0.100$ & $0.410$ & $0.386 \pm 0.044$ & $1244.0 \pm  17.0$ & $2.368 \pm 0.045$ & $1.110 \pm 0.039$ \\
HD128311  &    b & $5033 \pm 50$ & $ 0.05 \pm 0.08$ & $4.66 \pm 0.10$ & 2 & $0.345 \pm 0.049$ & $0.345$ & $1.457 \pm 0.152$ & $ 454.2 \pm   1.6$ & $1.086 \pm 0.018$ & $0.583 \pm 0.028$ \\
HD171238  &    b & $5557 \pm 50$ & $ 0.18 \pm 0.08$ & $4.55 \pm 0.10$ & 1 & $0.400 \pm 0.060$ & $0.400$ & $2.609 \pm 0.148$ & $1523.0 \pm  43.0$ & $2.543 \pm 0.064$ & $1.051 \pm 0.046$ \\
HD181433  &    c & $4914 \pm 50$ & $ 0.36 \pm 0.08$ & $4.35 \pm 0.10$ & 3 & $0.280 \pm 0.020$ & $0.480$ & $0.640 \pm 0.027$ & $ 962.0 \pm  15.0$ & $1.756 \pm 0.034$ & $1.008 \pm 0.070$ \\
HD181433  &    d & $4914 \pm 50$ & $ 0.36 \pm 0.08$ & $4.35 \pm 0.10$ & 3 & $0.480 \pm 0.050$ & $0.480$ & $0.535 \pm 0.051$ & $2172.0 \pm 158.0$ & $3.022 \pm 0.155$ & $1.008 \pm 0.070$ \\
HD190360  &    b & $5572 \pm 50$ & $ 0.21 \pm 0.08$ & $4.26 \pm 0.10$ & 2 & $0.313 \pm 0.019$ & $0.313$ & $1.535 \pm 0.061$ & $2915.0 \pm  28.9$ & $3.973 \pm 0.071$ & $1.075 \pm 0.037$ \\
HD202206  &    c & $5740 \pm 50$ & $ 0.24 \pm 0.08$ & $4.43 \pm 0.10$ & 2 & $0.267 \pm 0.021$ & $0.435$ & $2.331 \pm 0.127$ & $1383.4 \pm  18.4$ & $2.490 \pm 0.055$ & $0.986 \pm 0.035$ \\
HD204313  &    d & $5677 \pm 50$ & $ 0.11 \pm 0.08$ & $4.22 \pm 0.10$ & 2 & $0.280 \pm 0.090$ & $0.280$ & $1.606 \pm 0.281$ & $2831.6 \pm 150.0$ & $3.945 \pm 0.154$ & $0.000 \pm 0.000$ \\
HD207832  &    c & $5712 \pm 50$ & $ 0.10 \pm 0.08$ & $4.44 \pm 0.10$ & 2 & $0.270 \pm 0.160$ & $0.270$ & $0.730 \pm 0.161$ & $1155.7 \pm  54.5$ & $2.112 \pm 0.100$ & $0.901 \pm 0.056$ \\
HD217107  &    c & $5732 \pm 50$ & $ 0.37 \pm 0.08$ & $4.44 \pm 0.10$ & 2 & $0.517 \pm 0.033$ & $0.517$ & $2.615 \pm 0.150$ & $4270.0 \pm 220.0$ & $5.334 \pm 0.204$ & $1.500 \pm 0.030$ \\
HD220773  &    b & $5878 \pm 50$ & $-0.01 \pm 0.08$ & $4.10 \pm 0.10$ & 1 & $0.510 \pm 0.100$ & $0.510$ & $1.450 \pm 0.251$ & $3724.7 \pm 463.0$ & $4.943 \pm 0.418$ & $1.351 \pm 0.049$ \\
HD50499  &     b & $6008 \pm 50$ & $ 0.29 \pm 0.08$ & $4.29 \pm 0.10$ & 1 & $0.254 \pm 0.203$ & $0.254$ & $1.745 \pm 0.140$ & $2457.9 \pm  37.9$ & $3.872 \pm 0.076$ & $1.184 \pm 0.040$ \\
HD66428  &     b & $5749 \pm 50$ & $ 0.31 \pm 0.08$ & $4.41 \pm 0.10$ & 1 & $0.465 \pm 0.030$ & $0.465$ & $2.750 \pm 0.195$ & $1973.0 \pm  31.0$ & $3.143 \pm 0.070$ & $0.980 \pm 0.034$ \\
HD79498  &     b & $5786 \pm 50$ & $ 0.25 \pm 0.08$ & $4.46 \pm 0.10$ & 1 & $0.590 \pm 0.020$ & $0.590$ & $1.346 \pm 0.073$ & $1966.1 \pm  41.0$ & $3.133 \pm 0.068$ & $1.128 \pm 0.044$ \\
HD87883  &     b & $5043 \pm 50$ & $ 0.02 \pm 0.08$ & $4.61 \pm 0.10$ & 1 & $0.530 \pm 0.120$ & $0.530$ & $1.756 \pm 0.282$ & $2754.0 \pm  87.0$ & $3.576 \pm 0.096$ & $0.778 \pm 0.038$ \\
\multicolumn{12}{l}{\textbf{Eccentric cool Jupiters with expanded mass limits}} \\
HD106252  &    b & $5870 \pm 50$ & $-0.12 \pm 0.08$ & $4.34 \pm 0.10$ & 1 & $0.482 \pm 0.011$ & $0.482$ & $6.959 \pm 0.257$ & $1531.0 \pm   4.7$ & $2.611 \pm 0.044$ & $1.123 \pm 0.038$ \\
HD108341  &    b & $5225 \pm 50$ & $ 0.16 \pm 0.08$ & $4.59 \pm 0.10$ & 1 & $0.850 \pm 0.085$ & $0.850$ & $3.477 \pm 4.643$ & $1129.0 \pm   7.0$ & $2.007 \pm 0.034$ & $0.790 \pm 0.030$ \\
HD204941  &    b & $5126 \pm 50$ & $-0.13 \pm 0.08$ & $4.58 \pm 0.10$ & 1 & $0.370 \pm 0.080$ & $0.370$ & $0.267 \pm 0.035$ & $1733.0 \pm  74.0$ & $2.554 \pm 0.084$ & $0.850 \pm 0.037$ \\
HD73267  &     b & $5399 \pm 50$ & $ 0.10 \pm 0.08$ & $4.36 \pm 0.10$ & 1 & $0.256 \pm 0.009$ & $0.256$ & $3.063 \pm 0.105$ & $1260.0 \pm   7.0$ & $2.198 \pm 0.038$ & $1.166 \pm 0.037$ \\
HD74156  &     c & $5973 \pm 50$ & $ 0.00 \pm 0.08$ & $4.17 \pm 0.10$ & 2 & $0.380 \pm 0.020$ & $0.630$ & $8.247 \pm 0.357$ & $2520.0 \pm  15.0$ & $3.900 \pm 0.067$ & $1.345 \pm 0.044$ \\	
\multicolumn{12}{l}{\textbf{Hot-Jupiters}} \\
     HAT-P-1 &      b & $5895 \pm 50$ & $0.04 \pm 0.08$ & $4.26 \pm 0.10$ & \nodata & \nodata & \nodata & \nodata & \nodata & \nodata & \nodata \\
HAT-P-13 &      b & $5837 \pm 50$ & $0.50 \pm 0.08$ & $4.29 \pm 0.10$ & \nodata & \nodata & \nodata & \nodata & \nodata & \nodata & \nodata \\
HAT-P-15 &      b & $5674 \pm 50$ & $0.29 \pm 0.08$ & $4.45 \pm 0.10$ & \nodata & \nodata & \nodata & \nodata & \nodata & \nodata & \nodata \\
HAT-P-16 &      b & $6121 \pm 50$ & $0.05 \pm 0.08$ & $4.31 \pm 0.10$ & \nodata & \nodata & \nodata & \nodata & \nodata & \nodata & \nodata \\
HAT-P-17 &      b & $5377 \pm 50$ & $0.09 \pm 0.08$ & $4.58 \pm 0.10$ & \nodata & \nodata & \nodata & \nodata & \nodata & \nodata & \nodata \\
HAT-P-19 &      b & $5017 \pm 50$ & $0.34 \pm 0.08$ & $4.51 \pm 0.10$ & \nodata & \nodata & \nodata & \nodata & \nodata & \nodata & \nodata \\
HAT-P-20 &      b & $4653 \pm 50$ & $0.22 \pm 0.08$ & $4.61 \pm 0.10$ & \nodata & \nodata & \nodata & \nodata & \nodata & \nodata & \nodata \\
HAT-P-21 &      b & $5632 \pm 50$ & $0.04 \pm 0.08$ & $4.30 \pm 0.10$ & \nodata & \nodata & \nodata & \nodata & \nodata & \nodata & \nodata \\
HAT-P-22 &      b & $5355 \pm 50$ & $0.26 \pm 0.08$ & $4.39 \pm 0.10$ & \nodata & \nodata & \nodata & \nodata & \nodata & \nodata & \nodata \\
HAT-P-23 &      b & $5900 \pm 50$ & $0.09 \pm 0.08$ & $4.45 \pm 0.10$ & \nodata & \nodata & \nodata & \nodata & \nodata & \nodata & \nodata \\
HAT-P-25 &      b & $5604 \pm 51$ & $0.43 \pm 0.08$ & $4.48 \pm 0.10$ & \nodata & \nodata & \nodata & \nodata & \nodata & \nodata & \nodata \\
HAT-P-27 &      b & $5346 \pm 50$ & $0.31 \pm 0.08$ & $4.52 \pm 0.10$ & \nodata & \nodata & \nodata & \nodata & \nodata & \nodata & \nodata \\
HAT-P-29 &      b & $6061 \pm 50$ & $0.11 \pm 0.08$ & $4.27 \pm 0.10$ & \nodata & \nodata & \nodata & \nodata & \nodata & \nodata & \nodata \\
HAT-P-3 &      b & $5209 \pm 50$ & $0.35 \pm 0.08$ & $4.54 \pm 0.10$ & \nodata & \nodata & \nodata & \nodata & \nodata & \nodata & \nodata \\
HAT-P-4 &      b & $6070 \pm 50$ & $0.33 \pm 0.08$ & $4.31 \pm 0.10$ & \nodata & \nodata & \nodata & \nodata & \nodata & \nodata & \nodata \\
HAT-P-44 &      b & $5372 \pm 50$ & $0.37 \pm 0.08$ & $4.43 \pm 0.10$ & \nodata & \nodata & \nodata & \nodata & \nodata & \nodata & \nodata \\
HAT-P-46 &      b & $6197 \pm 50$ & $0.31 \pm 0.08$ & $4.30 \pm 0.10$ & \nodata & \nodata & \nodata & \nodata & \nodata & \nodata & \nodata \\
K00001   &     01 & $5870 \pm 50$ & $0.05 \pm 0.08$ & $4.47 \pm 0.10$ & \nodata & \nodata & \nodata & \nodata & \nodata & \nodata & \nodata \\
K00017   &     01 & $5775 \pm 50$ & $0.48 \pm 0.08$ & $4.41 \pm 0.10$ & \nodata & \nodata & \nodata & \nodata & \nodata & \nodata & \nodata \\
K00022   &     01 & $5850 \pm 50$ & $0.16 \pm 0.08$ & $4.29 \pm 0.10$ & \nodata & \nodata & \nodata & \nodata & \nodata & \nodata & \nodata \\
K00094   &     01 & $6176 \pm 50$ & $0.01 \pm 0.08$ & $4.23 \pm 0.10$ & \nodata & \nodata & \nodata & \nodata & \nodata & \nodata & \nodata \\
K00098   &     01 & $6482 \pm 50$ & $0.01 \pm 0.08$ & $4.14 \pm 0.10$ & \nodata & \nodata & \nodata & \nodata & \nodata & \nodata & \nodata \\
K00127   &     01 & $5611 \pm 50$ & $0.34 \pm 0.08$ & $4.50 \pm 0.10$ & \nodata & \nodata & \nodata & \nodata & \nodata & \nodata & \nodata \\
K00128   &     01 & $5679 \pm 50$ & $0.34 \pm 0.08$ & $4.32 \pm 0.10$ & \nodata & \nodata & \nodata & \nodata & \nodata & \nodata & \nodata \\
K00135   &     01 & $6012 \pm 50$ & $0.35 \pm 0.08$ & $4.37 \pm 0.10$ & \nodata & \nodata & \nodata & \nodata & \nodata & \nodata & \nodata \\
K00203   &     01 & $5331 \pm 50$ & $0.50 \pm 0.08$ & $4.28 \pm 0.12$ & \nodata & \nodata & \nodata & \nodata & \nodata & \nodata & \nodata \\
K00351   &     01 & $5976 \pm 50$ & $0.11 \pm 0.08$ & $4.32 \pm 0.10$ & \nodata & \nodata & \nodata & \nodata & \nodata & \nodata & \nodata \\
K00433   &     02 & $5233 \pm 50$ & $0.29 \pm 0.08$ & $4.54 \pm 0.10$ & \nodata & \nodata & \nodata & \nodata & \nodata & \nodata & \nodata \\
K00682   &     01 & $5566 \pm 50$ & $0.34 \pm 0.08$ & $4.27 \pm 0.10$ & \nodata & \nodata & \nodata & \nodata & \nodata & \nodata & \nodata \\
K00686   &     01 & $5593 \pm 50$ & $0.06 \pm 0.09$ & $4.42 \pm 0.15$ & \nodata & \nodata & \nodata & \nodata & \nodata & \nodata & \nodata \\
K01411   &     01 & $5718 \pm 50$ & $0.41 \pm 0.08$ & $4.35 \pm 0.10$ & \nodata & \nodata & \nodata & \nodata & \nodata & \nodata & \nodata \\
K03681   &     01 & $6258 \pm 50$ & $0.16 \pm 0.08$ & $4.39 \pm 0.10$ & \nodata & \nodata & \nodata & \nodata & \nodata & \nodata & \nodata \\
\enddata
	\tablecomments{Stellar parameters ($\teff, \mh, \logg$) were derived using SPC. The parameters for the hot-Jupiter Kepler KOIs originate from \citep{buchhave_three_2014}. All remaining parameters are taken from exoplanets.org (17 June, 2016).}
\end{deluxetable*}


\begin{thebibliography}{}
	\expandafter\ifx\csname natexlab\endcsname\relax\def\natexlab#1{#1}\fi
	
	\bibitem[{Alexander \& Pascucci(2012)}]{alexander_deserts_2012}
	Alexander, R.~D., \& Pascucci, I. 2012, Monthly Notices of the Royal
	Astronomical Society, 422, L82
	
	\bibitem[{Bai \& Stone(2010)}]{bai_effect_2010}
	Bai, X.-N., \& Stone, J.~M. 2010, The Astrophysical Journal Letters, 722, L220
	
	\bibitem[{Baruteau {et~al.}(2014)Baruteau, Crida, Paardekooper, Masset, Guilet,
		Bitsch, Nelson, Kley, \& Papaloizou}]{baruteau_planet-disk_2014}
	Baruteau, C., Crida, A., Paardekooper, S.-J., {et~al.} 2014, Protostars and
	Planets VI, 667
	
	\bibitem[{Bitsch {et~al.}(2015{\natexlab{a}})Bitsch, Johansen, Lambrechts, \&
		Morbidelli}]{bitsch_structure_2015}
	Bitsch, B., Johansen, A., Lambrechts, M., \& Morbidelli, A. 2015{\natexlab{a}},
	Astronomy and Astrophysics, 575, A28
	
	\bibitem[{Bitsch {et~al.}(2015{\natexlab{b}})Bitsch, Lambrechts, \&
		Johansen}]{bitsch_growth_2015}
	Bitsch, B., Lambrechts, M., \& Johansen, A. 2015{\natexlab{b}}, Astronomy and
	Astrophysics, 582, A112
	
	\bibitem[{Borucki {et~al.}(2010)Borucki, Koch, Basri, Batalha, Brown, Caldwell,
		Caldwell, Christensen-Dalsgaard, Cochran, DeVore, Dunham, Dupree, Gautier,
		Geary, Gilliland, Gould, Howell, Jenkins, Kondo, Latham, Marcy, Meibom,
		Kjeldsen, Lissauer, Monet, Morrison, Sasselov, Tarter, Boss, Brownlee, Owen,
		Buzasi, Charbonneau, Doyle, Fortney, Ford, Holman, Seager, Steffen, Welsh,
		Rowe, Anderson, Buchhave, Ciardi, Walkowicz, Sherry, Horch, Isaacson,
		Everett, Fischer, Torres, Johnson, Endl, MacQueen, Bryson, Dotson, Haas,
		Kolodziejczak, Van~Cleve, Chandrasekaran, Twicken, Quintana, Clarke, Allen,
		Li, Wu, Tenenbaum, Verner, Bruhweiler, Barnes, \& Prsa}]{borucki_kepler_2010}
	Borucki, W.~J., Koch, D., Basri, G., {et~al.} 2010, Science, 327, 977
	
	\bibitem[{Brauer {et~al.}(2007)Brauer, Dullemond, Johansen, Henning, Klahr, \&
		Natta}]{brauer_survival_2007}
	Brauer, F., Dullemond, C.~P., Johansen, A., {et~al.} 2007, Astronomy and
	Astrophysics, 469, 1169
	
	\bibitem[{Bryan {et~al.}(2016)Bryan, Knutson, Howard, Ngo, Batygin, Crepp,
		Fulton, Hinkley, Isaacson, Johnson, Marcy, \& Wright}]{bryan_statistics_2016}
	Bryan, M.~L., Knutson, H.~A., Howard, A.~W., {et~al.} 2016, The Astrophysical
	Journal, 821, 89
	
	\bibitem[{Buchhave {et~al.}(2012)Buchhave, Latham, Johansen, Bizzarro, Torres,
		Rowe, Batalha, Borucki, Brugamyer, Caldwell, Bryson, Ciardi, Cochran, Endl,
		Esquerdo, Ford, Geary, Gilliland, Hansen, Isaacson, Laird, Lucas, Marcy,
		Morse, Robertson, Shporer, Stefanik, Still, \&
		Quinn}]{buchhave_abundance_2012}
	Buchhave, L.~A., Latham, D.~W., Johansen, A., {et~al.} 2012, Nature, 486, 375
	
	\bibitem[{Buchhave {et~al.}(2014)Buchhave, Bizzarro, Latham, Sasselov, Cochran,
		Endl, Isaacson, Juncher, \& Marcy}]{buchhave_three_2014}
	Buchhave, L.~A., Bizzarro, M., Latham, D.~W., {et~al.} 2014, Nature, 509, 593
	
	\bibitem[{Chatterjee {et~al.}(2008)Chatterjee, Ford, Matsumura, \&
		Rasio}]{chatterjee_dynamical_2008}
	Chatterjee, S., Ford, E.~B., Matsumura, S., \& Rasio, F.~A. 2008, The
	Astrophysical Journal, 686, 580
	
	\bibitem[{Dawson \& Murray-Clay(2013)}]{dawson_giant_2013}
	Dawson, R.~I., \& Murray-Clay, R.~A. 2013, The Astrophysical Journal Letters,
	767, L24
	
	\bibitem[{Ercolano \& Clarke(2010)}]{ercolano_metallicity_2010}
	Ercolano, B., \& Clarke, C.~J. 2010, Monthly Notices of the Royal Astronomical
	Society, 402, 2735
	
	\bibitem[{Fischer \& Valenti(2005)}]{fischer_planet-metallicity_2005}
	Fischer, D.~A., \& Valenti, J. 2005, The Astrophysical Journal, 622, 1102
	
	\bibitem[{Fűrész(2008)}]{furesz_design_2008}
	Fűrész, G. 2008, PhD thesis, University of Szeged, Szeged, Hungary
	
	\bibitem[{Huber {et~al.}(2014)Huber, Silva~Aguirre, Matthews, Pinsonneault,
		Gaidos, García, Hekker, Mathur, Mosser, Torres, Bastien, Basu, Bedding,
		Chaplin, Demory, Fleming, Guo, Mann, Rowe, Serenelli, Smith, \&
		Stello}]{huber_revised_2014}
	Huber, D., Silva~Aguirre, V., Matthews, J.~M., {et~al.} 2014, The Astrophysical
	Journal Supplement Series, 211, 2
	
	\bibitem[{Ida {et~al.}(2016)Ida, Guillot, \& Morbidelli}]{ida_radial_2016}
	Ida, S., Guillot, T., \& Morbidelli, A. 2016, Astronomy and Astrophysics, 591,
	A72
	
	\bibitem[{Kooten {et~al.}(2016)Kooten, Wielandt, Schiller, Nagashima, Thomen,
		Larsen, Olsen, Nordlund, Krot, \& Bizzarro}]{kooten_isotopic_2016}
	Kooten, E. M. M. E.~V., Wielandt, D., Schiller, M., {et~al.} 2016, Proceedings
	of the National Academy of Sciences, 113, 2011
	
	\bibitem[{Lambrechts \& Johansen(2012)}]{lambrechts_rapid_2012}
	Lambrechts, M., \& Johansen, A. 2012, Astronomy and Astrophysics, 544, A32
	
	\bibitem[{Lambrechts \& Johansen(2014)}]{lambrechts_forming_2014}
	---. 2014, Astronomy and Astrophysics, 572, A107
	
	\bibitem[{Lambrechts {et~al.}(2014)Lambrechts, Johansen, \&
		Morbidelli}]{lambrechts_separating_2014}
	Lambrechts, M., Johansen, A., \& Morbidelli, A. 2014, Astronomy and
	Astrophysics, 572, A35
	
	\bibitem[{Lega {et~al.}(2013)Lega, Morbidelli, \& Nesvorný}]{lega_early_2013}
	Lega, E., Morbidelli, A., \& Nesvorný, D. 2013, Monthly Notices of the Royal
	Astronomical Society, 431, 3494
	
	\bibitem[{Levison {et~al.}(2015)Levison, Kretke, \&
		Duncan}]{levison_growing_2015}
	Levison, H.~F., Kretke, K.~A., \& Duncan, M.~J. 2015, Nature, 524, 322
	
	\bibitem[{Mamajek(2009)}]{mamajek_initial_2009}
	Mamajek, E.~E. 2009, in , eprint: arXiv:0906.5011, 3--10
	
	\bibitem[{Mayor {et~al.}(2003)Mayor, Pepe, Queloz, Bouchy, Rupprecht, Lo~Curto,
		Avila, Benz, Bertaux, Bonfils, Dall, Dekker, Delabre, Eckert, Fleury,
		Gilliotte, Gojak, Guzman, Kohler, Lizon, Longinotti, Lovis, Megevand,
		Pasquini, Reyes, Sivan, Sosnowska, Soto, Udry, van Kesteren, Weber, \&
		Weilenmann}]{mayor_setting_2003}
	Mayor, M., Pepe, F., Queloz, D., {et~al.} 2003, The Messenger, 114, 20
	
	\bibitem[{Mayor {et~al.}(2011)Mayor, Marmier, Lovis, Udry, Ségransan, Pepe,
		Benz, Bertaux, Bouchy, Dumusque, Curto, Mordasini, Queloz, \&
		Santos}]{mayor_harps_2011}
	Mayor, M., Marmier, M., Lovis, C., {et~al.} 2011, arXiv:1109.2497 [astro-ph],
	arXiv: 1109.2497
	
	\bibitem[{Morbidelli {et~al.}(2015)Morbidelli, Lambrechts, Jacobson, \&
		Bitsch}]{morbidelli_great_2015}
	Morbidelli, A., Lambrechts, M., Jacobson, S., \& Bitsch, B. 2015, Icarus, 258,
	418
	
	\bibitem[{Morbidelli {et~al.}(2005)Morbidelli, Levison, Tsiganis, \&
		Gomes}]{morbidelli_chaotic_2005}
	Morbidelli, A., Levison, H.~F., Tsiganis, K., \& Gomes, R. 2005, Nature, 435,
	462
	
	\bibitem[{Morbidelli {et~al.}(2012)Morbidelli, Lunine, O'Brien, Raymond, \&
		Walsh}]{morbidelli_building_2012}
	Morbidelli, A., Lunine, J.~I., O'Brien, D.~P., Raymond, S.~N., \& Walsh, K.~J.
	2012, Annual Review of Earth and Planetary Sciences, 40, 251
	
	\bibitem[{Morbidelli {et~al.}(2016)Morbidelli, Bitsch, Crida, Gounelle,
		Guillot, Jacobson, Johansen, Lambrechts, \&
		Lega}]{morbidelli_fossilized_2016}
	Morbidelli, A., Bitsch, B., Crida, A., {et~al.} 2016, Icarus, 267, 368
	
	\bibitem[{Mulders {et~al.}(2016)Mulders, Pascucci, Apai, Frasca, \&
		Molenda-Żakowicz}]{mulders_super-solar_2016}
	Mulders, G.~D., Pascucci, I., Apai, D., Frasca, A., \& Molenda-Żakowicz, J.
	2016, The Astronomical Journal, 152, 187
	
	\bibitem[{Naef {et~al.}(2010)Naef, Mayor, Lo~Curto, Bouchy, Lovis, Moutou,
		Benz, Pepe, Queloz, Santos, Ségransan, Udry, Bonfils, Delfosse, Forveille,
		Hébrard, Mordasini, Perrier, Boisse, \& Sosnowska}]{naef_harps_2010}
	Naef, D., Mayor, M., Lo~Curto, G., {et~al.} 2010, Astronomy \& Astrophysics,
	523, A15
	
	\bibitem[{Ndugu {et~al.}(2018)Ndugu, Bitsch, \& Jurua}]{ndugu_planet_2018}
	Ndugu, N., Bitsch, B., \& Jurua, E. 2018, Monthly Notices of the Royal
	Astronomical Society, 474, 886
	
	\bibitem[{Ormel \& Klahr(2010)}]{ormel_effect_2010}
	Ormel, C.~W., \& Klahr, H.~H. 2010, Astronomy and Astrophysics, 520, A43
	
	\bibitem[{Paardekooper {et~al.}(2011)Paardekooper, Baruteau, \&
		Meru}]{paardekooper_numerical_2011}
	Paardekooper, S.-J., Baruteau, C., \& Meru, F. 2011, Monthly Notices of the
	Royal Astronomical Society, 416, L65
	
	\bibitem[{Petigura {et~al.}(2017)Petigura, Howard, Marcy, Johnson, Isaacson,
		Cargile, Hebb, Fulton, Weiss, Morton, Winn, Rogers, Sinukoff, Hirsch, \&
		Crossfield}]{petigura_california-kepler_2017}
	Petigura, E.~A., Howard, A.~W., Marcy, G.~W., {et~al.} 2017, The Astronomical
	Journal, 154, 107
	
	\bibitem[{Pollack {et~al.}(1996)Pollack, Hubickyj, Bodenheimer, Lissauer,
		Podolak, \& Greenzweig}]{pollack_formation_1996}
	Pollack, J.~B., Hubickyj, O., Bodenheimer, P., {et~al.} 1996, Icarus, 124, 62
	
	\bibitem[{Santos {et~al.}(2004)Santos, Israelian, \&
		Mayor}]{santos_spectroscopic_2004}
	Santos, N.~C., Israelian, G., \& Mayor, M. 2004, Astronomy and Astrophysics,
	415, 1153
	
	\bibitem[{Santos {et~al.}(2013)Santos, Sousa, Mortier, Neves, Adibekyan,
		Tsantaki, Delgado~Mena, Bonfils, Israelian, Mayor, \&
		Udry}]{santos_sweet-cat:_2013}
	Santos, N.~C., Sousa, S.~G., Mortier, A., {et~al.} 2013, Astronomy and
	Astrophysics, 556, A150
	
	\bibitem[{Shabram {et~al.}(2016)Shabram, Demory, Cisewski, Ford, \&
		Rogers}]{shabram_eccentricity_2016}
	Shabram, M., Demory, B.-O., Cisewski, J., Ford, E.~B., \& Rogers, L. 2016, The
	Astrophysical Journal, 820, 93
	
	\bibitem[{Sousa {et~al.}(2011)Sousa, Santos, Israelian, Mayor, \&
		Udry}]{sousa_spectroscopic_2011}
	Sousa, S.~G., Santos, N.~C., Israelian, G., Mayor, M., \& Udry, S. 2011,
	Astronomy and Astrophysics, 533, 141
	
	\bibitem[{Telting {et~al.}(2014)Telting, Avila, Buchhave, Frandsen, Gandolfi,
		Lindberg, Stempels, {the NOT staff}, \& Prins}]{telting_fies:_2014}
	Telting, J., Avila, G., Buchhave, L., {et~al.} 2014, Astronomische Nachrichten,
	335, 41
	
	\bibitem[{Tsiganis {et~al.}(2005)Tsiganis, Gomes, Morbidelli, \&
		Levison}]{tsiganis_origin_2005}
	Tsiganis, K., Gomes, R., Morbidelli, A., \& Levison, H.~F. 2005, Nature, 435,
	459
	
	\bibitem[{Vogt {et~al.}(1994)Vogt, Allen, Bigelow, Bresee, Brown, Cantrall,
		Conrad, Couture, Delaney, Epps, Hilyard, Hilyard, Horn, Jern, Kanto, Keane,
		Kibrick, Lewis, Osborne, Pardeilhan, Pfister, Ricketts, Robinson, Stover,
		Tucker, Ward, \& Wei}]{vogt_hires:_1994}
	Vogt, S.~S., Allen, S.~L., Bigelow, B.~C., {et~al.} 1994, in  (International
	Society for Optics and Photonics), 362--376
	
	\bibitem[{Wang {et~al.}(2016)Wang, Wang, Wu, Zhao, Li, Luo, Liu, Zhang, Hou,
		Wang, \& Cao}]{wang_calibration_2016}
	Wang, L., Wang, W., Wu, Y., {et~al.} 2016, The Astronomical Journal, 152, 6
	
	\bibitem[{Yasui {et~al.}(2009)Yasui, Kobayashi, Tokunaga, Saito, \&
		Tokoku}]{yasui_lifetime_2009}
	Yasui, C., Kobayashi, N., Tokunaga, A.~T., Saito, M., \& Tokoku, C. 2009, The
	Astrophysical Journal, 705, 54
	
\end{thebibliography}
\end{document}